%% file: conference_101719.tex
\documentclass[conference]{IEEEtran}
\IEEEoverridecommandlockouts
\usepackage{amsmath,amssymb,amsfonts}
\usepackage{algorithmic}
\usepackage{graphicx}
\usepackage{textcomp}
\usepackage[dvipsnames]{xcolor}
\usepackage{subcaption}

\usepackage{amssymb}
\usepackage{amsmath}
\usepackage{booktabs}
\usepackage{bm}
\usepackage{bbm}
\usepackage{multirow}
\usepackage{url}
\usepackage{pgfplots}
\usepgfplotslibrary{fillbetween}
\usetikzlibrary{arrows}
\usetikzlibrary{backgrounds}
\usetikzlibrary{decorations.pathreplacing}

\newcommand{\bd}{\boldsymbol}
\definecolor{mygreen}{RGB}{250, 166, 65}

\begin{document}

\title{Learning Mobility Flows from Urban Features with Spatial Interaction Models and Neural Networks\textsuperscript{*}
}
% \footnote{To appear in the Proceedings of 2020 IEEE International Conference on Smart Computing (SMARTCOMP 2020)}

\author{
\thanks{*To appear in the Proceedings of 2020 IEEE International Conference on Smart Computing (SMARTCOMP 2020)}

\IEEEauthorblockN{Gevorg Yeghikyan}
\IEEEauthorblockA{
\textit{Scuola Normale Superiore}\\
Pisa, Italy \\
gevorg.yeghikyan@sns.it
}
\and
\IEEEauthorblockN{Felix L. Opolka}
\IEEEauthorblockA{\textit{University of Cambridge}\\
Cambridge, UK \\
flo23@cam.ac.uk
}
\and
\IEEEauthorblockN{Mirco Nanni}
\IEEEauthorblockA{
\textit{ISTI-CNR}\\
Pisa, Italy \\
mirco.nanni@isti.cnr.it
}
\and
\IEEEauthorblockN{Bruno Lepri}
\IEEEauthorblockA{
\textit{FBK}\\
Trento, Italy\\
lepri@fbk.eu
}
\and
\IEEEauthorblockN{Pietro Li\`o}
\IEEEauthorblockA{\textit{University of Cambridge}\\
Cambridge, UK\\
pl219@cam.ac.uk}
}

\maketitle

\begin{abstract}
A fundamental problem of interest to policy makers, urban planners, and other stakeholders involved in urban development projects is assessing the impact of planning and construction activities on mobility flows. This is a challenging task due to the different spatial, temporal, social, and economic factors influencing urban mobility flows. These flows, along with the influencing factors, can be modelled as attributed graphs with both node and edge features characterising locations in a city and the various types of relationships between them. In this paper, we address the problem of assessing origin-destination (OD) car flows between a location of interest and every other location in a city, given their features and the structural characteristics of the graph. We propose three neural network architectures, including graph neural networks (GNN), and conduct a systematic comparison between the proposed methods and state-of-the-art spatial interaction models, their modifications, and machine learning approaches. The objective of the paper is to address the practical problem of estimating potential flow between an urban development project location and other locations in the city, where the features of the project location are known in advance. We evaluate the performance of the models on a regression task using a custom data set of attributed car OD flows in London. 
%  Second, we show that despite the recent successes of GNNs on different types of networks across many fields, they do not seem to outperform simpler conventional neural network architectures on the formulated task on mobility network data.
We also visualise the model performance by showing the spatial distribution of flow residuals across London.
\end{abstract}


\section{Introduction} 
Planning and managing city and transportation infrastructures requires understanding the relationship between urban mobility flows and spatial, structural, and socio-economic features associated with them.
There exists extensive literature addressing this problem ranging from the classical gravity model and its modifications~\cite{wilson1971family,fotheringham1989spatial} to the more recent spatial econometric interaction models~\cite{lesage2008spatial} and the non-parametric radiation models~\cite{simini2012universal} that attempt to characterise cross-sectional origin-destination (OD) flow matrices. Furthermore, various neural network-based models have been proposed for predicting temporal OD flow matrices~\cite{chai2018bike,toque2016forecasting}.
However, modelling OD flow matrices in their entirety, the mentioned works do not address the problem of assessing flows between a specific location and every other location in the city, given all other flows, other location characteristics, as well as information on the dyadic relations between those locations.

More specifically, the motivation for this task comes from a scenario in which it is necessary to assess the impact of an urban development project on the OD flows in and out of the project's location. Examples of these 
motivating scenarios include retail location choice and consumer spatial behaviour prediction, which have been approached with the Huff model and its modifications~\cite{huff1963probabilistic}. These models, however, suffer from a series of drawbacks related mostly to overly restrictive assumptions. 
In this paper, we take a different approach and focus on the problem of evaluating OD flows in and out of a location of interest. By modelling urban flows as attributed graphs in which the nodes represent locations in a city (i.e. each node is described by a vector of features such as population density, Airbnb prices, available parking areas, etc.), and the edges represent the car flows between them (each one described by a vector of features such as road distance, average time required to travel, average speed, etc.), this project aims to offer an instrument for assessing flows between a specific location and all other locations in the city.

Since a rigorous experimental setting would have required difficult-to-obtain longitudinal data of OD flows \textit{before} and \textit{after} the completion of an urban development project, we set up a \textit{quasi-experimental} setting.
We randomly select locations in a city and the flows associated with them as a test set, and attempt to find a function that takes the urban features describing city locations and the remaining flows as input, and predicts the flows in the test set as output. 

In sum, our paper makes the following contributions:
\begin{itemize}
\item We propose three neural network architectures for predicting car flows between a location of interest and every other location in a city. Two of the models use graph convolutional layers that pool information from geographical or topological neighbourhoods around relevant nodes to incorporate more information (Section~\ref{sec:methodology}).
\item We evaluate and compare our models on a custom dataset of aggregate OD car flows in London, containing node and edge features (Section~\ref{sec:experiments}).
\item We show that the proposed neural network models outperform well-known spatial interaction and machine learning models. A comparison among neural network models reveals that graph convolutions do not substantially improve prediction performance on the formulated task (Sections~\ref{sec:problem}, \ref{sec:experiments}).
\item We describe our custom dataset and make it publicly available along with the code for this study (Section~\ref{sec:data}).
\end{itemize}

% In particular, in the proposed end-to-end Graph Convolutional Neural Network framework we follow a geographically embedded spatial graph-based approach, which allows the model to learn the intricate relationships between urban features and OD flows by pooling information from both geographical and topological neighbourhoods of the location (node) of interest.
% We evaluate and compare all the models on a custom dataset of aggregate OD car flows in London, built with features describing city locations and dyadic relationships between these locations. The evaluation on the regression task of predicting the magnitude of flow to (from) a location of interest show that the neural network based models achieve, among other measures, a Mean Absolute Error (\textit{MAE}) of an order of magnitude lower than most classical spatial interaction models. We show, however, that contrary to our expectations, the novel Graph Convolutional Neural Network model, which pools geographical and topological neighbourhood information in the learning process, does not outperform the simpler fully connected Neural Network model agnostic to the graph structure and information. 

% The rest of the paper is structured as follows: The next section presents the related work, followed by the problem statement, the detailed description of the dataset we release with the paper, and the proposed methodology. Next, we discuss the experiments, the results and finally we draw some conclusions.

\begin{figure*}  
  \begin{subfigure}[b]{0.32\textwidth}
    \includegraphics[width=\linewidth]{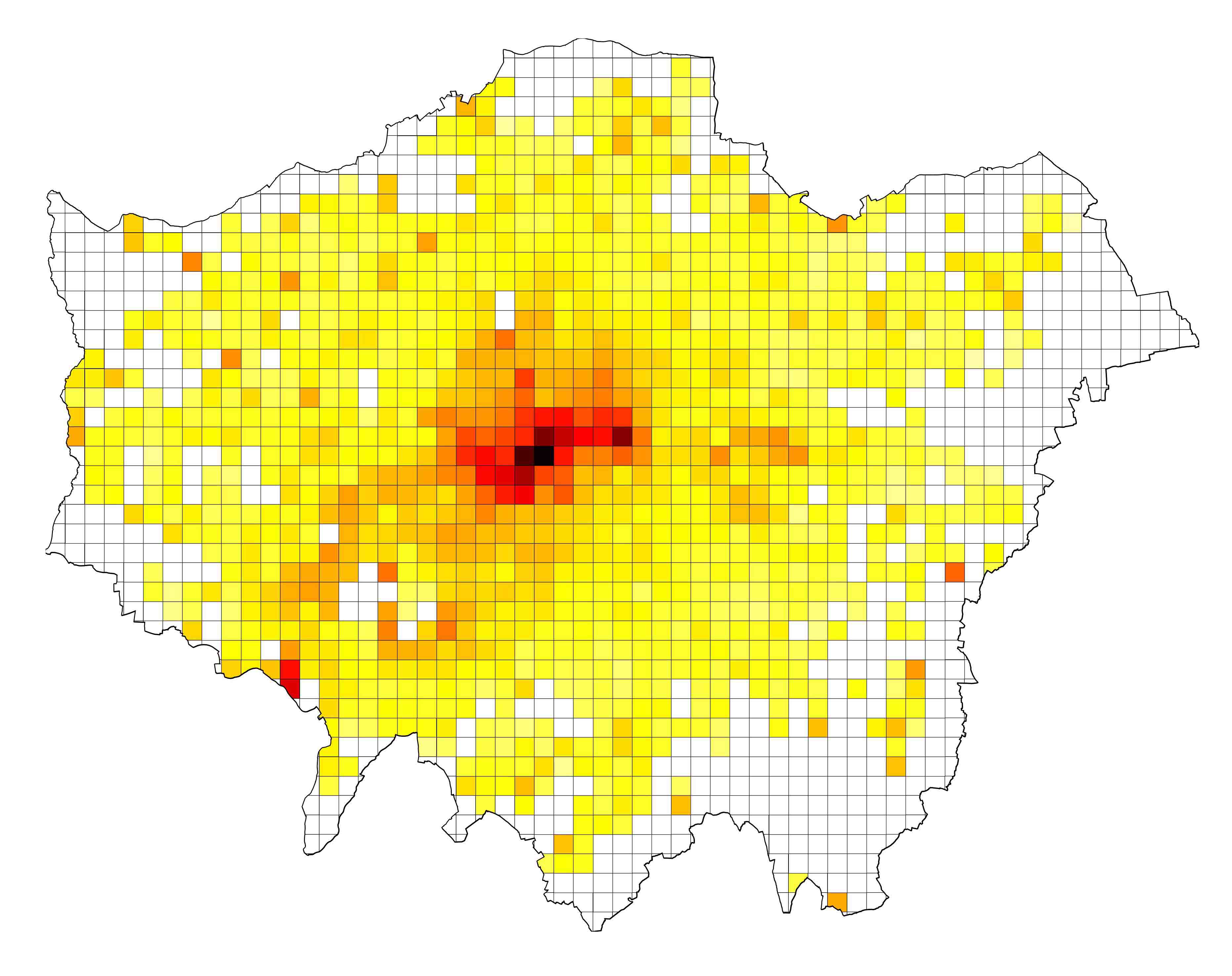}
    \caption{}
  \end{subfigure}
  \hfill
  \begin{subfigure}[b]{0.32\textwidth}
    \includegraphics[width=\linewidth]{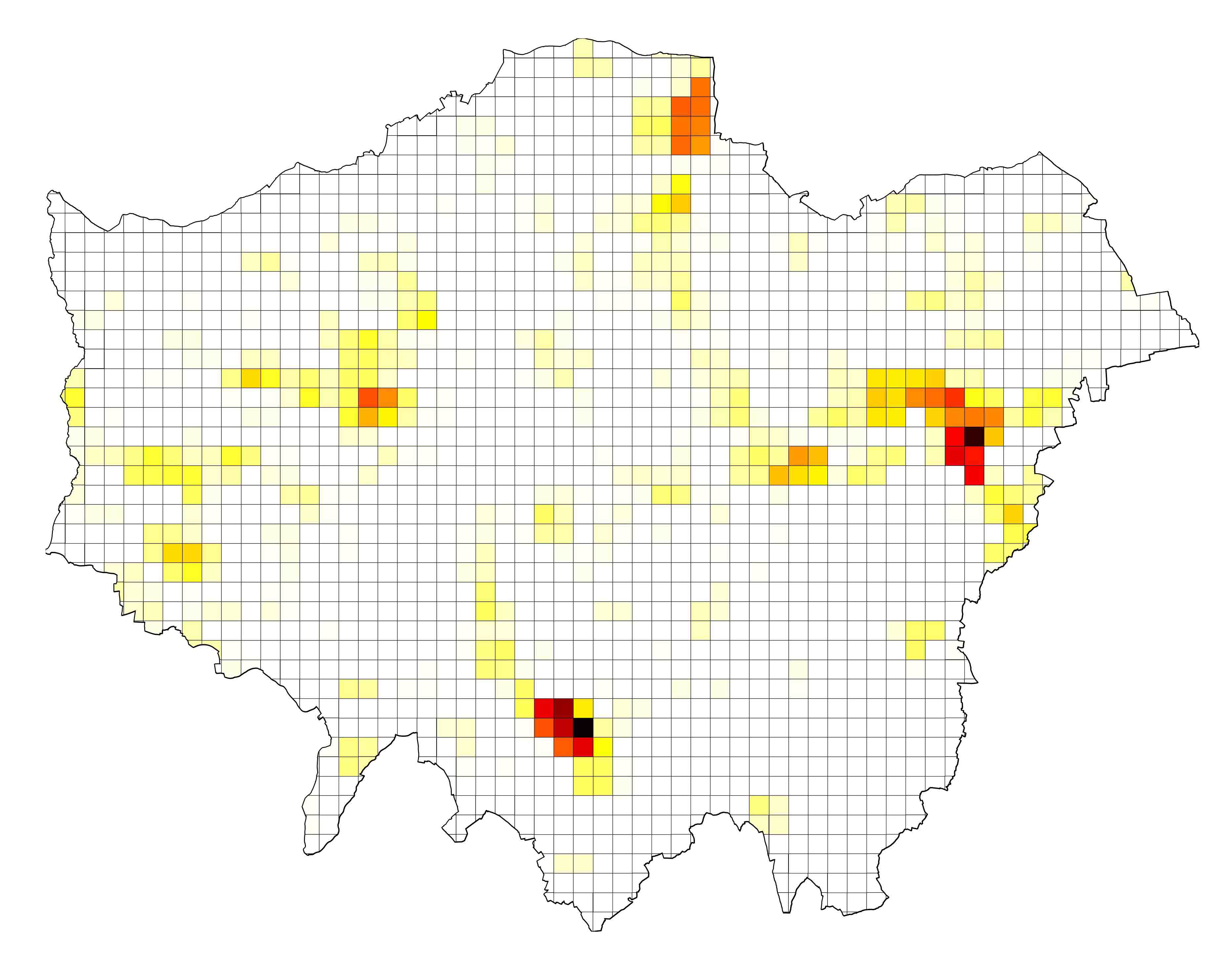}
    \caption{}
  \end{subfigure}
  \hfill
  \begin{subfigure}[b]{0.32\textwidth}
    \includegraphics[width=\linewidth]{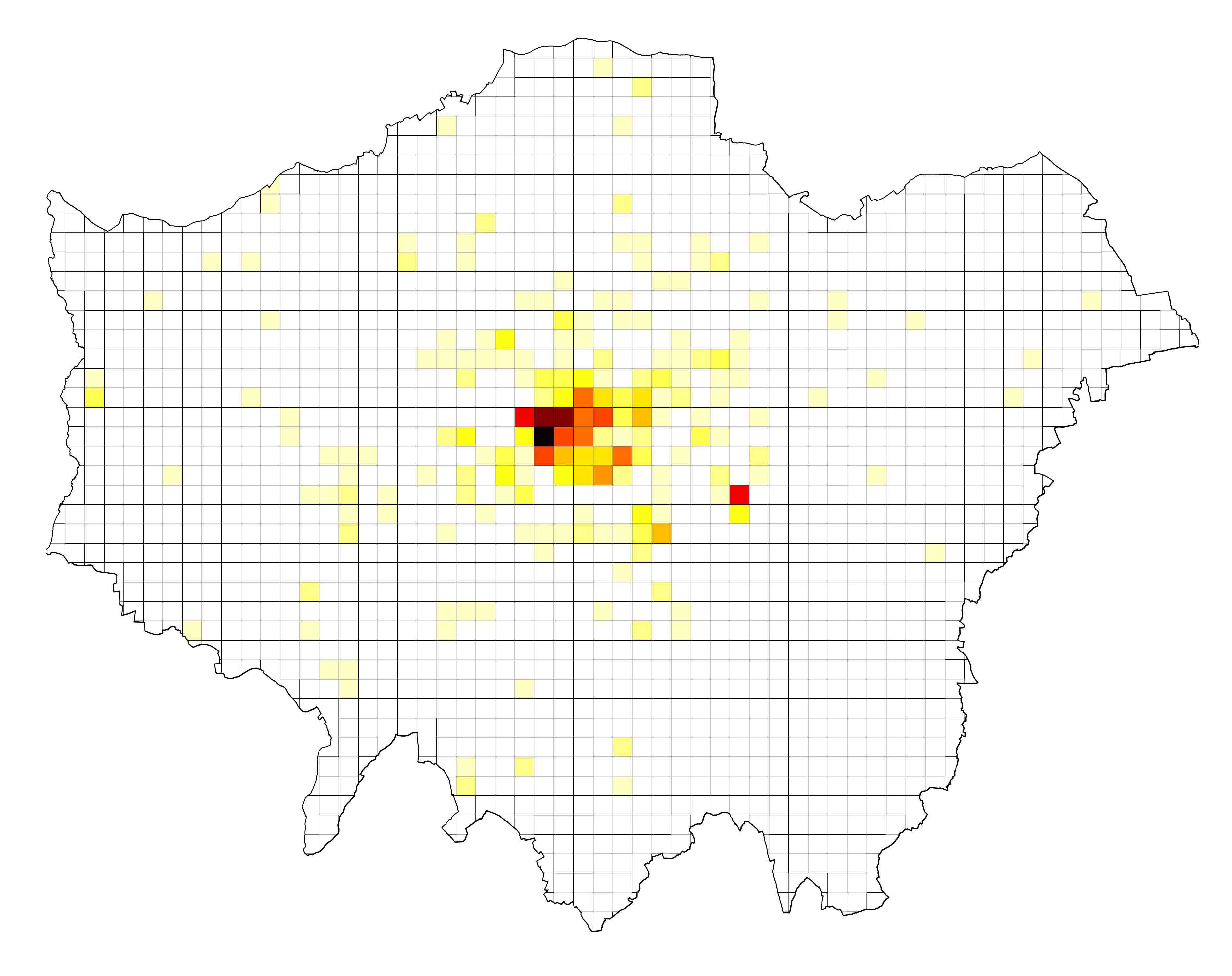}
    \caption{}
  \end{subfigure}
  \caption{Examples of node (cell) features (a) Average Airbnb listing prices (b) Proportion of grid cell area allotted to industrial activity (c) Number of museums and galleries per grid cell. Darker colours indicate higher values.}
  \label{fig:node_attrs}
\end{figure*}

\section{Related work} \label{related_work}

The problem of estimating human flows between locations in a geographical space has been first addressed by~\cite{wilson1971family} through a family of spatial interaction models and subsequently extended by~\cite{fotheringham1989spatial}.
Spatial interaction models, extensively used to estimate human mobility flows and trip demand between locations as a function of the location features, have become an acknowledged method for modelling geographical mobility in transportation planning~\cite{erlander1990gravity,de2011modelling}, commuting~\cite{mcarthur2011spatial}, and spatial economics~\cite{patuelli2007network}. The spatial interaction models are usually calibrated via an Ordinary Least Squares (OLS) regression, which assumes normally distributed data. However, OD flows are usually not distributed normally, are count data, and contain a large number of zero flows. This makes the setting incompatible with OLS estimation and requires either a Poisson model or, in the presence of over-dispersion, a Negative Binomial Regression (NB) model~\cite{zhang2019spatial}.

Another major concern in this modelling scenario are the complex interactions often caused by spatial dependencies and non-stationarity. The former arises from spill-over effects from a location to its neighbourhoods, while the latter is caused by the influence of independent variables varying across space. These issues have been addressed in literature by spatial autocorrelation and geographically weighted modelling techniques~\cite{fotheringham2003geographically,lesage2008spatial,da2014geographically,zhang2019spatial}.

Another approach within the spatial interaction modelling paradigm is the Huff model and its extensions~\cite{huff1963probabilistic}. Originally developed mainly for retail location choice and turnover prediction, they represent a probabilistic formulation of the gravity model. The Huff model considers OD flows as proportional to the relative attractiveness and accessibility of the destination compared to other competing destinations. The probability $P_{ij}$ of a consumer at location $i$ of choosing to shop at a retail location $j$ is framed as:
    \begin{equation}
        P_{i j}=\frac{A_{j}^{\alpha} D_{i j}^{-\beta}}{\sum_{j=1}^{n} A_{j}^{\alpha} D_{i j}^{-\beta}},
    \end{equation}
where $A_{j}$ is a measure of attractiveness of retail location $j$, such as area or a linear combination of different features, $D_{ij}$ is the distance between locations $i$ and $j$, $\alpha$ and $\beta$, estimated from empirical observations, are attractiveness and distance decay parameters, respectively.

Along with traditional gravity methods, the Huff model and its variations have found their way to numerous applications including location selection of movie theaters~\cite{davis2006spatial}, a university campus~\cite{bruno2008using}, or the analysis of spatial access to health care~\cite{wan2012three}.

However, these models suffer from too restrictive assumptions such as considering the ratio of the probabilities of an individual selecting two alternatives as being unaffected by the introduction of a third alternative. Although the competing destinations model~\cite{fotheringham1983new} has overcome this, it has the disadvantage of considering either spatial agglomeration or competition effects, ignoring the fact that they can coexist in the same location. Even though a number of extensions to the Huff model and the gravity framework in general have been proposed to overcome spatial non-stationarity and to include a larger array of features affecting the flows~\cite{de2014extended,li2012assessing}, this family of models, along with the non-parametric radiation and population-weighted opportunities model, have demonstrated to fall short of high predictive capacity particularly at the city scale~\cite{masucci2013gravity,liang2013unraveling,yan2014universal}.

More recently, machine learning, particularly a Random Forest approach, has shown promising results in reconstructing inter-city OD flow matrices~\cite{spadon2019reconstructing}. However, its performance on intra-urban flow data remains to be tested.

Moreover, as already mentioned, the discussed models address the problem of modelling the OD flow matrix as a whole and have to be adapted to our specific task of estimating flows between a specific location and all other locations, given the other flows in the city, the location features, and the features describing the dyadic relations between them, respectively.

The problem of estimating OD flows has also been addressed with neural network methods~\cite{lorenzo2013od}. As flows are most naturally modelled by graphs, most work has focused on the use of graph neural networks for flow estimation.
An early neural network model for graph structured data has been suggested in~\cite{scarselli2009ggn}. Later work has specifically focused on generalising Convolutional Neural Networks from the domain of regular grids to the domain of irregular graphs~\cite{bruna2014gcn, defferrard2016chebnet}. One of the most commonly used graph neural network models is the Graph Convolutional Neural Network (GCN) proposed in~\cite{kipf2017gcn}. 

Graph neural networks have previously been applied to urban planning tasks. In~\cite{chai2018bike}, they have been used to predict the flow of bikes within a bike sharing system. Unlike our model, flows are modelled as node-level features, which requires a different neural network model and does not allow to predict flows between specific pairs of nodes.
Although~\cite{wang2019cellular} uses graph neural networks to predict flows between parts of a city, their model operates on spatio-temporal data and focuses on the temporal aspect of the data.
Beyond flow prediction, in~\cite{zhu2018patterns}, a graph neural network model has been proposed for building site selection. 
A broader overview of machine learning methods applied to the task of urban flow prediction is given in~\cite{xie2019survey}.
In this work, we define neural network models that make use of stationary node and edge features and compare different neural network architectures based on fully connected networks and graph neural networks. 

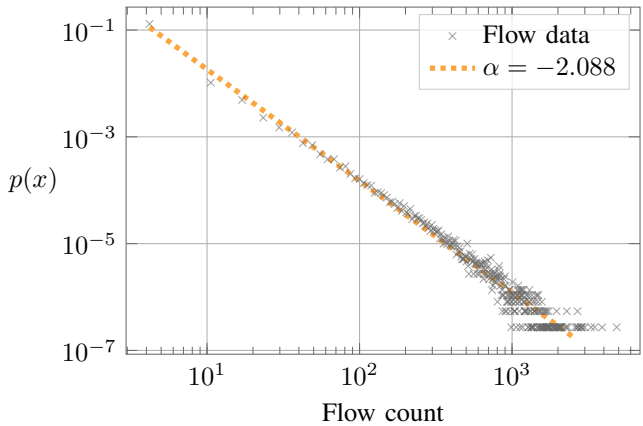
\begin{figure}
    \centering
    \input{loglogplot}
    \caption{Log-log plot of the probability distribution of the OD flows fitted with a power-law distribution $p(x) \propto x^{-\alpha}$ with exponent of $\alpha = -2.088$.}
    \label{fig:flowdist}
\end{figure}

\section{Data description}\label{sec:data}
\begin{figure*}
  \begin{subfigure}[b]{0.32\textwidth}
    \includegraphics[width=\linewidth]{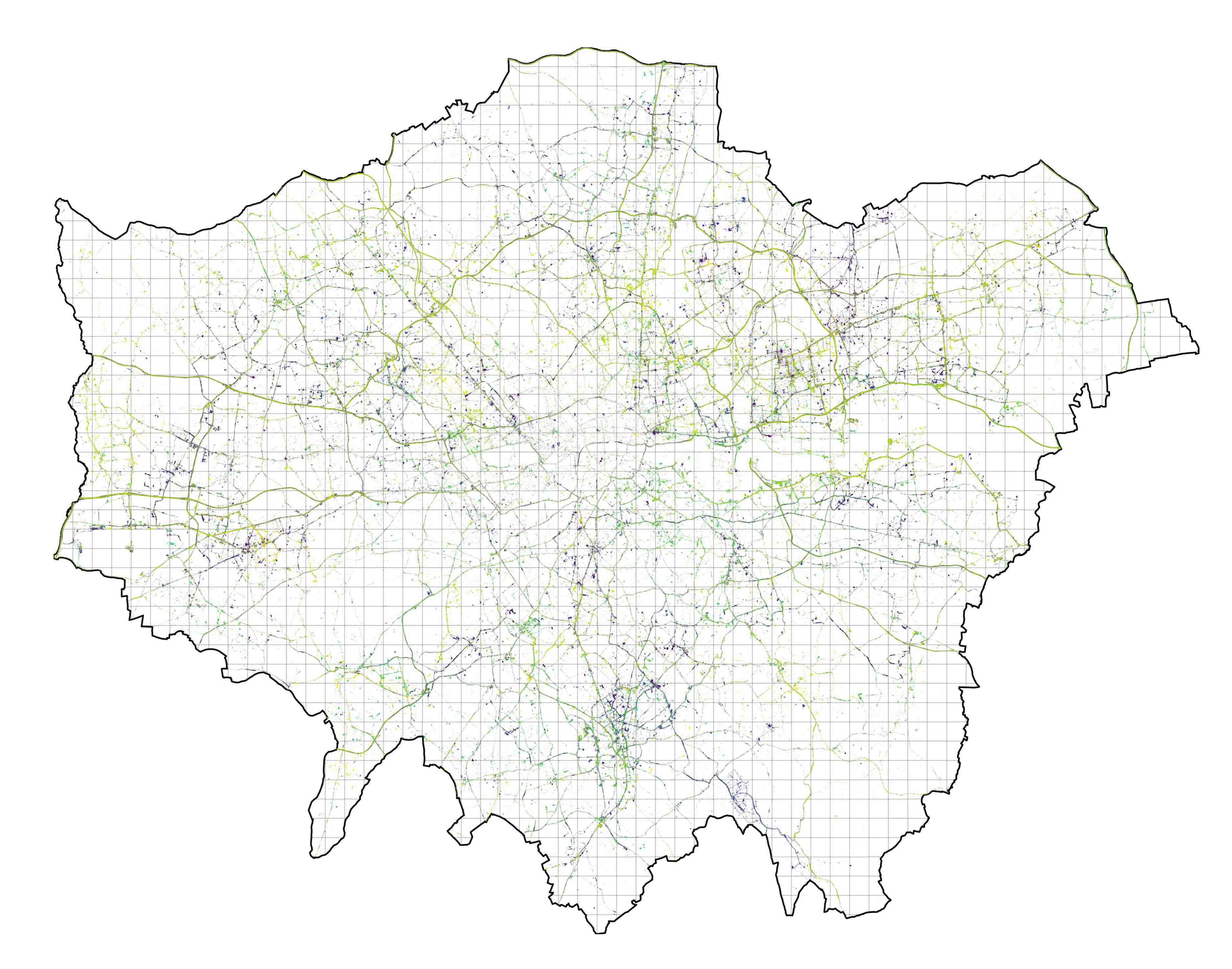}
    \caption{}
    \label{London_grid_gps}
  \end{subfigure}
  \hfill
  \begin{subfigure}[b]{0.32\textwidth}
    \includegraphics[width=\linewidth]{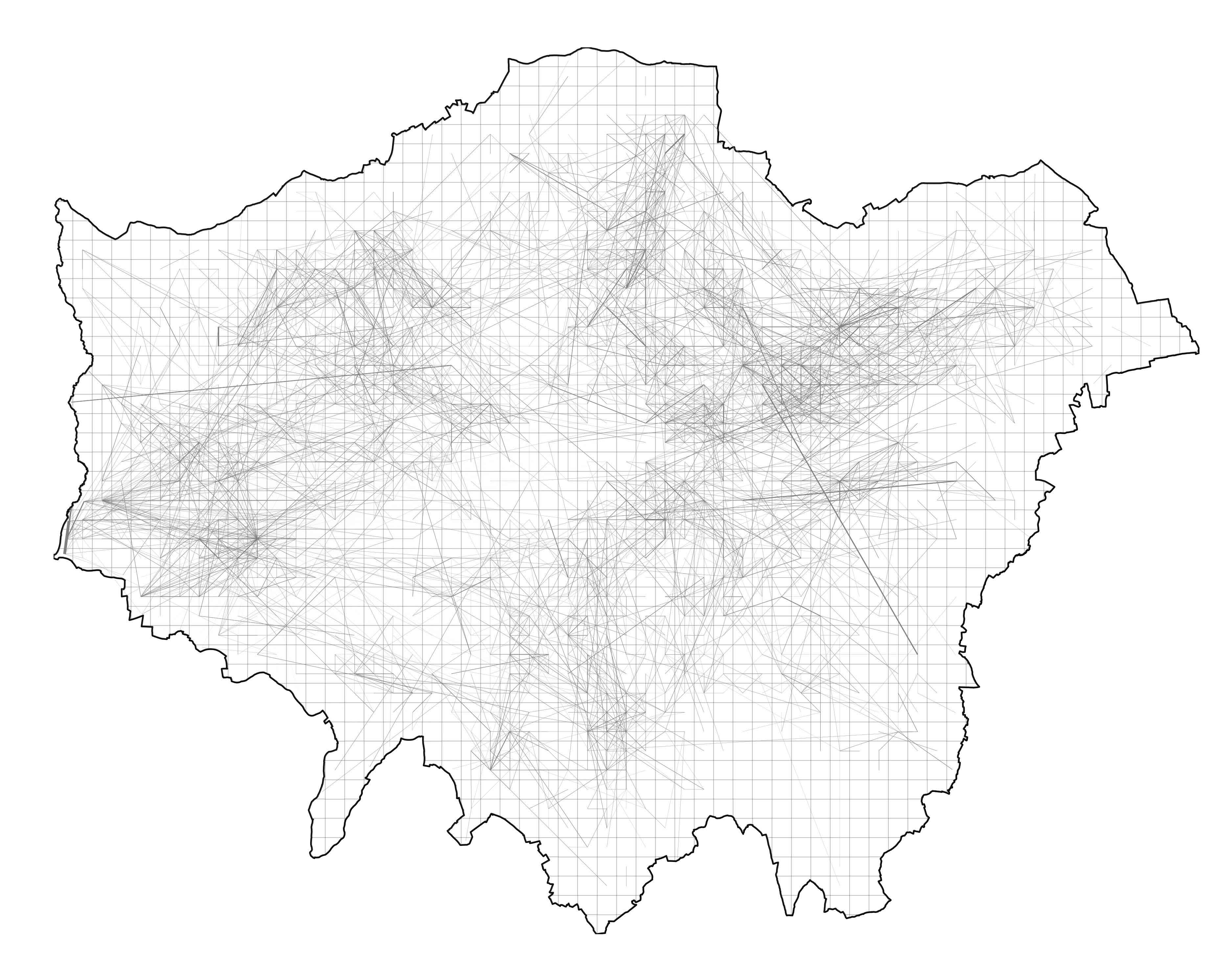}
    \caption{}
    \label{London_flows_gps}
  \end{subfigure}
  \hfill
  \begin{subfigure}[b]{0.32\textwidth}
    \includegraphics[width=\linewidth]{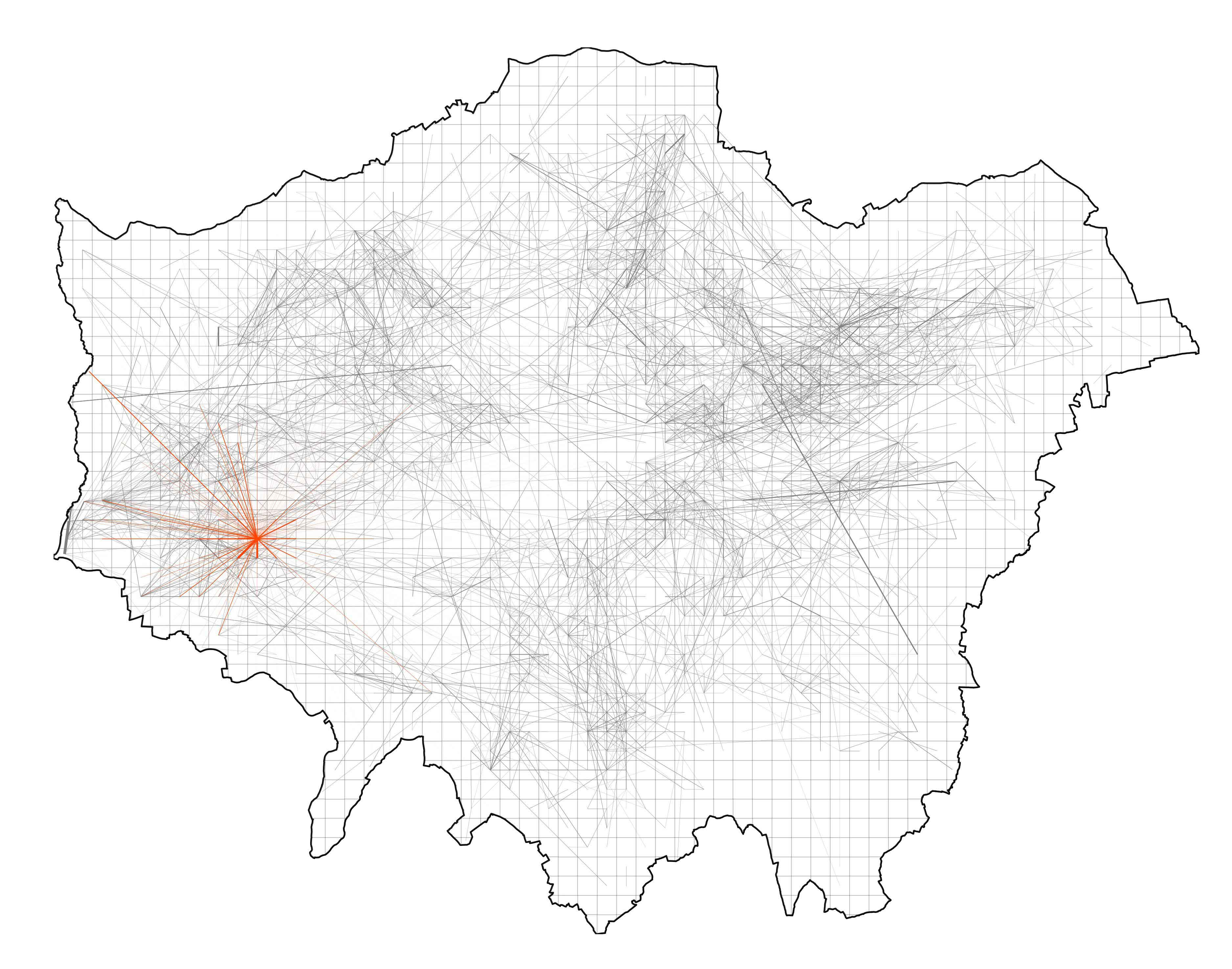}
    \caption{}
    \label{London_flow_network2}
  \end{subfigure}
  \caption{(a) Car GPS trajectories over grid cells in London. (b) Origin-Destination ($OD$) flow network in London. (c) Target flows between a node of interest and every other node.}
\end{figure*}

We publicly release\footnote{Dataset will be released at https://trackandknowproject.eu/file-repository/. Code available at \url{github.com/FelixOpolka/Mobility-Flows-Neural-Networks}.} a custom dataset of aggregate origin-destination (OD) flows of private cars in London augmented with feature data describing city locations and dyadic relations between them. 
The workflow of building the dataset is as follows:
\begin{enumerate}
    \item The urban territory has been subdivided into $n$ Cartesian grid cells of size $500 \times 500$ m, and each such quadratic cell is considered a node in the graph.
    \item The GPS trajectories of around 10000 cars spanning a period of one year, provided by a car insurance company for research purposes, have been superimposed on the grid, and trip origins and destinations have been extracted (Figure~\ref{London_grid_gps}).
    \item The OD network has been built from the extracted origin-destination pairs by aggregating the flow counts over a year (Figure~\ref{London_flows_gps}).
    % described by the weighted adjacency matrix $\textbf{W} \in \mathbb{R}^{n \times n}$, where $W_{ij}$ represents the number of car trips between nodes $i$ and $j$. 
    Since the aggregation spans such a long time period, the OD matrix is approximately symmetric, and thus has been converted into a symmetric matrix by averaging the matrix with its transpose.
    \item The node features have been built by engineering 35 features from various open sources \cite{OpenStreetMap, Airbnb, LondonTransport} and from the GPS data. These features include population density, average Airbnb prices, parking areas, areas covered by residential buildings, number of restaurants, bars, banks, museums, road network density, average radius of gyration, etc. per cell. Examples of node features and their spatial distribution are visualised in Figure~\ref{fig:node_attrs}.
    \item Similarly, the edge features encode information on 12 dyadic relations such as network distance, average time, average speed, temporal correlation between car incidence in cells, public transport connections, etc. The detailed attribute description is provided with the dataset.  
\end{enumerate}

\section{Problem statement} \label{sec:problem}
In this section, we describe the problem we are addressing and state definitions of important terms.

We define a \textbf{weighted attributed graph} $G = (\mathcal{V}, \mathcal{E}, \textbf{W}, \textbf{X}^{v}, \textbf{X}^{e})$ with feature information associated with both nodes and edges. More specifically, $\mathcal{V}$ is the set of $n$ nodes, and $\mathcal{E}=\left\{e_{ij} = (i,j) : i,j \in \mathcal{V}\right\}$ represents the set of $m$ edges in graph $G$. Furthermore, $\textbf{W} \in \mathbb{R}^{n \times n}$ is the weighted adjacency matrix, essentially the OD matrix, with $\textbf{W}_{ij} \geq 0 \; \forall i,j \in \mathcal{V}$ corresponding to the flow between cells $i$ and $j$. Additionally, we denote the node feature matrix as $\textbf{X}^{v} \in \mathbb{R}^{n \times p}$, where $p$ is the number of node features. The edge feature matrix, on the other hand, is denoted as $\textbf{X}^{e} \in \mathbb{R}^{m \times k}$, where $k$ is the number of edge features.

The urban mobility flow network $T$ is a weighted undirected attributed graph whose nodes are $500 \times 500$ m city grid cells, and the edges are the aggregate flows between them. The nodes and edges are additionally augmented by feature vectors described in detail in Section~\ref{sec:data}.
Furthermore, each edge $e_{ij}$ in the urban mobility flow network $T$ is associated with a target (or ground truth) flow $w_{ij}$, which is the corresponding entry in the weighted adjacency matrix $\bd{W}$ of $T$. It represents the aggregate mobility flow between cell (node) $i$ and cell (node) $j$ in the network.

In our prediction setting, we are given the urban mobility flow network $T = (\mathcal{V}, \mathcal{E}, \textbf{W}, \textbf{X}^{v}, \textbf{X}^{e})$ and a node of interest $i$ for which the target flows $W_{i1}, \ldots, W_{in}$ are unknown. Hence, we aim to learn a mapping $f :\left\{\mathcal{V}, \mathcal{E}, \textbf{W}, \textbf{X}^{v}, \textbf{X}^{e}\right\} \rightarrow \mathbb{R}^{n}$ from the urban mobility flow network to the missing flows, i.e. $[W_{i1}, \ldots, W_{in}] = f(i, \textbf{W}, \textbf{X}^{v}, \textbf{X}^{e}) \; \forall i \in \mathcal{V}$. In other words, the aim is to predict the missing target flows (Figure~\ref{London_flow_network2}), given the features of node $i$ and the rest of the graph.

\section{Methodology}\label{sec:methodology}

In the following, we describe three neural network models that are trained to predict the unknown flows in the urban mobility flow network $T$. When a model makes a prediction for the flow associated with an edge going from a node of interest to another node in the graph, it can use all node and edge features in the graph, as these features are available even for nodes of interest, i.e. sites of prospective urban development projects. Furthermore, it may use the ground truth flows for edges that are not connected to a node of interest. In a practical situation, this corresponds to the flows between \textit{existing} locations in the city for which flow information is therefore available.

The first neural network architecture is a fully connected neural network operating on the features of the target edge and the features of its two incident nodes.
More specifically, when predicting the flow for target edge $e_{ij}$, we concatenate the node features $\bd{x}_i^v$ and $\bd{x}_j^v$ for incident node features, as well as the corresponding edge features $\bd{x}_{ij}^e$.
The concatenated vector
    \begin{equation}
        \bd{\bar{x}} = [\bd{x}_i^v, \bd{x}_{ij}^e, \bd{x}_j^v]\label{eq:concat}
    \end{equation}
is passed into a fully connected neural network with ReLU-non-linearities, defined as $\text{ReLU}(z_j) = \text{max}(0, z_j)$, where $z_j$ is the $j^\text{th}$ output of the linear transformation.
Each fully connected layer is followed by batch normalisation~\cite{ioffe2015batch} and dropout~\cite{srivastava2014dropout} to counter overfitting. We refer to this model as \textit{FCNN}.

\begin{figure*}
    \centering
    \input{model_architecture}
    \caption{Overview of the neural network model architectures. When predicting the flow for edge $e_{ij}$, all three models concatenate the corresponding edge features $\bd{x}_{ij}^e$, and the node features $\bd{x}_i^v, \bd{x}_j^v$ of the incident nodes. The resulting vector is fed into a single fully connected layer. In case of the GNN-based models \textit{GNN-geo} and \textit{GNN-flow}, the network also perform graph convolutions on the neighbourhoods of $v_i$ and $v_j$ and computes a weighted sum of both neighbourhood embeddings and the edge embedding. A further set of fully connected layers maps the sum to the predicted flow $\hat{y}_{ij}$. The \textit{FCNN} model skips the addition step and does not perform graph convolutions.}
    \label{fig:overview}
\end{figure*}
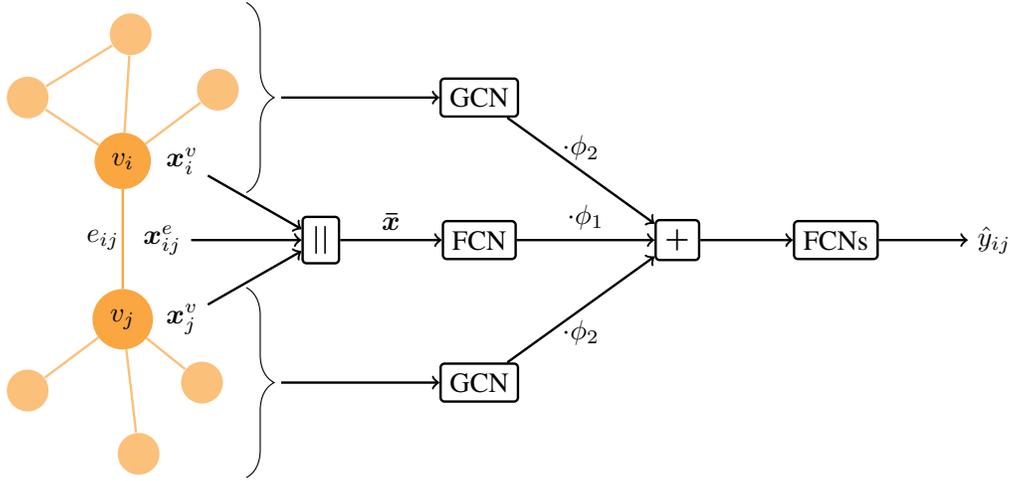

The second model builds upon the \textit{FCNN} model through the additional use of graph convolutions to generate embeddings of node neighbourhoods. We use a graph convolutional neural network (GCN)~\cite{kipf2017gcn} to generate node embeddings $\bd{h}_i, \bd{h}_j$ for the two nodes incident to the target edge $e_{ij}$. 
GCN layers extend fully-connected layers with an additional neighbourhood aggregation step before the non-linearity. The layer applies a linear transformation to all node features $\bd{h}_i^{(l-1)}$ in the graph and then, for each node, computes a weighted average of the resulting representations at the central node and in the 1-hop neighbourhood of the central node:
    \begin{equation}
        \bd{z}_i^{(l)} = \sum_{j \in \mathcal{N}(i) \cup \{i \}} \frac{1}{\sqrt{(d_i + 1)(d_j + 1)}} \bd{h}_j^{(l-1)} \bd{\Theta},
    \end{equation}
where $\bd{\Theta} \in \mathbb{R}^{D^{(l-1)} \times D^{(l)}}$ is a learned weight matrix, $\mathcal{N}(i)$ refers to the 1-hop neighbourhood of node $i$, and $d_i$ denotes the degree of node $i$. This aggregation scheme is followed by a non-linearity and can be written more compactly using matrix multiplication as
    \begin{equation}
        \bd{H}^{(l)} = \text{ReLU}(\bd{\tilde{D}}^{-\frac{1}{2}} \bd{\tilde{W}} \bd{\tilde{D}}^{-\frac{1}{2}} \bd{H}^{(l-1)} \bd{\Theta}).\label{eq:gcn}
    \end{equation}
where $\bd{\tilde{W}} = \bd{W} + \bd{I}$ and $\bd{\tilde{D}}$ is the degree matrix of $\bd{\tilde{W}}$.
Equation~\ref{eq:gcn} defines a graph convolutional layer and multiple such layers can be stacked to form a multi-layer graph neural network.
A GNN with $k$ layers allows us to compute embeddings encoding node feature information from within a $k$-hop neighbourhood.

For the second model, we apply multiple graph convolutions as defined above on the flow-weighted geographical adjacency matrix $\bd{W}^{\text{geo}}$ where $W_{ij}^{\text{geo}}$ is non-zero if and only if node $i$ is in the geographical neighbourhood of node $j$ and $W_{ij}^{\text{geo}} = W_{ij}$, i.e. the flow between $i$ and $j$. The resulting node embeddings $\bd{h}_i, \bd{h}_j \in \mathbb{R}^{D}$ for the two nodes incident to edge $e_{ij}$ are added to the representation of $\bd{\bar{x}}$ (see Equation~\ref{eq:concat} after the first fully connected layer:
    \begin{equation}
        \bd{h}_{ij}^{(1)} = \phi_1 \text{FCN}(\bd{\bar{x}}) + \phi_2 \left[ \text{GNN}(\bd{x}_i) + \text{GNN}(\bd{x}_j) \right],
    \end{equation}
where $\phi_1, \phi_2$ are learned weighting coefficients.
We note that both mentions of $\text{GNN}(\cdot)$ refer to the same sequence of graph convolutional layers. We then feed $\bd{h}_{ij}^{(1)}$ into a number of fully connected layers, again with dropout and batch normalisation, such that the resulting model contains the same number of fully connected layers as the \textit{FCNN} model. We call the resulting model \textit{GNN-geo}.

Finally, we evaluate a third model, denoted by \textit{GNN-flow}, which is equivalent to \textit{GNN-geo} except graph convolutions are performed using the flow-based adjacency matrix $\bd{W}^{\text{flow}} = \bd{W}$, where $W_{ij}^{flow}$ is the flow between $i$ and $j$. Hence, the adjacency matrix used by \textit{GNN-flow} will contain additional edges to those used by \textit{GNN-geo}. A visualisation of the model architectures is given in Figure~\ref{fig:overview}.

The graph based models \textit{GNN-geo} and \textit{GNN-flow} require flow information for the adjacency matrices. While this is readily available for edges between two regular nodes, we have to approximate flow between a regular node $i$ and a node of interest $j$. This is done by taking the average of the flows from node $i$ to each node in the neighbourhood of $j$, i.e.
\begin{equation}
    \tilde{W}_{ij} = \frac{1}{\vert \mathcal{N}(j) \vert} \sum\limits_{k \in \mathcal{N}(j)} W_{ik}.\label{eq:approx_flows}
\end{equation}

We note that even though the \textit{FCNN} does not use graph convolutions and hence does not qualify as a common graph neural network, it does use graph structure information by concatenating specifically the features $\bd{x}_i^v, \bd{x}_j^v$ of the nodes incident to the target edge $e_{ij}$.

All models output the flow corresponding to the target edge $e_{ij}$ and are trained to minimise the mean squared error between the predicted and the actual flow. More details on the experimental setup are provided in Section~\ref{sec:exp_setup}.

\section{Experiments}\label{sec:experiments}

We evaluate the described model on the London dataset described in Section~\ref{sec:data}. In the following, we describe the goodness-of-fit metrics we use to measure model performance, the baseline methods we compare our models to, and the experimental setup.

\subsection{Goodness-of-fit measures}

\textit{Mean absolute error (MAE).} Let $\hat{y}_{ij}$ be the predicted flow between $i$ and $j$, $y_{ij}$ be the ground truth flow, then

    \begin{equation}
        \mathrm{MAE}=\frac{1}{|\mathcal{E}|} \sum_{i} \sum_{j} \left|y_{ij}-\hat{y}_{ij}\right|.
    \end{equation}

\textit{Binned MAE.} Due to the highly skewed distribution of the flow data, the vast majority of flows have a small flow count, with only a handful of flows with a very large flow value (see Figure~\ref{fig:flowdist}). Because of this, the total MAE will be biased downwards. To account for this, we additionally measure the MAE of all models within 4 bins with the following boundaries: 0 $\leq$ 10.0 $\leq$ 100.0 $\leq$ 1000.0 $\leq$ 10000.0, corresponding to $\text{MAE}_0$, $\text{MAE}_1$, $\text{MAE}_2$, $\text{MAE}_3$, respectively.
Finally, we define the MAE bin mean as
    \begin{equation}
        \text{Bin mean MAE} = \frac{\text{MAE}_{0} + \text{MAE}_{1} + \text{MAE}_{2} + \text{MAE}_{3}}{4}\label{eq:binmean},
    \end{equation}
where $\text{MAE}_i$ refers to MAE of the $i^{\text{th}}$ bin.

% \begin{figure}[tbp]
% \centering
% \includegraphics[width=0.97\columnwidth]{London_flow_dist.jpg}
% \caption{Log-log plot of the probability density function of the OD flows with a fitted power-law exponent of $\alpha = -2.088$.}
% \label{flow_dist}
% \end{figure}

\textit{Mean absolute percentage error (MAPE).} To display the model accuracy with respect to the ground-truth flow values, we further use the mean absolute percentage error, defined as 

\begin{equation}
\mathrm{MAPE}=100 \times \frac{1}{|\mathcal{E}|} \sum_{i} \sum_{j} \left|\frac{y_{ij}-\hat{y}_{ij}}{y_{ij}}\right| ,
\end{equation}

\textit{Sorensen similarity index.} \ We use a modified version of the Sorensen similarity index (SSI), which has been extensively used in spatial interaction modelling \cite{yan2014universal, lenormand2012universal}, and is defined as
    \begin{equation} \label{eq:11}
        S S I=\frac{1}{|\mathcal{E}|} \sum_{i} \sum_{j} \frac{2 \min \left(y_{i j}, \hat{y}_{i j}\right)}{y_{i j}+\hat{y}_{ij}},
    \end{equation}
 and takes on values between 0 and 1, with values closer to 1 denoting a better fit.

\textit{Common part of commuters.} \ Further, we use a similar metric, the common part of commuters, used specifically for mobility OD flow networks \cite{lenormand2012universal}:
    \begin{equation} \label{eq:12}
        C P C=\frac{2 \sum_{i, j=1}^{n} \min \left(y_{i j}, \hat{y}_{i j}\right)}{\sum_{i, j=1}^{n} y_{i j}+\sum_{i, j=1}^{n} \hat{y}_{i j}}.
    \end{equation}
This measure takes on the value 0, when the flows in the two networks completely differ, and 1, when they are in perfect agreement.

\textit{Common part of links.} \ Finally, to measure the degree to which the topological structure of the original network has been reconstructed, we use the common part of links (CPL) \cite{lenormand2016systematic} defined as
    \begin{equation} \label{eq:13}
        C P L=\frac{2 \sum_{i, j=1}^{n} \mathbbm{1}_{y_{i j}>0} \cdot \mathbbm{1}_{\hat{y}_{i j}>0}}{\sum_{i, j=1}^{n} \mathbbm{1}_{y_{i j}>0}+\sum_{i, j=1}^{n} \mathbbm{1}_{\hat{y}_{i j}>0}},
    \end{equation}
where $\mathbbm{1}_{A}$ is the indicator function of condition $A$. The common part of links
shows the proportion of links between the observed and predicted networks such that $y_{ij} > 0$ and $\hat{y}_{ij} > 0$. It takes on the value zero if the two networks have no common links and one if the networks are topologically equivalent.

% \textit{GEH statistic}  \ Finally, we borrow the classical GEH statistic from traffic engineering [cite] for comparing two sets of traffic volumes.

% \begin{equation}
% G E H (y_{ij}, \hat{y}_{ij}) =\sqrt{\frac{2(y_{ij}-\hat{y}_{ij})^{2}}{y_{ij}+\hat{y}_{ij}}},
% \end{equation}

% As empirical formula, it has been extensively used in traffic modelling. A GEH value of less than 5 is considered a good result and an accepted heuristic is obtaining at least 85\% of traffic volumes to have a GEH value less than 5.

\subsection{Baseline models}
In this study, we compare the proposed model to the following baselines, using the same experimental setup for all models:

\begin{itemize}
    \item \textbf{Doubly constrained gravity model (DC-GM)}: The classical gravity model with a power law decay has several formulations with respect to preserving the total in- nor out-flows during model calibration: unconstrained, origin-constrained, destination-constrained, and doubly constrained. Here we take the latter.
    \item \textbf{Huff model}: A probabilistic formulation of the gravity model described in Section \ref{related_work}.
    \item \textbf{Poisson regression}: An instance of the Generalized Linear Modelling framework, in which the dependent variable, being count data, is assumed to be drawn from a Poisson distribution. 
    \item \textbf{Negative Binomial regression (NB)}: A generalization of the Poisson regression in which the restrictive assumption that the mean and the variance of the dependent variable are equal is loosened.
    \item \textbf{Spatial Autoregressive Model (SAM)}: An extension to the Generalized Linear Modelling framework by accounting for spatial dependence among the flows by using spatial lags represented by spatial weight matrices built from observed data~\cite{lesage2008spatial}.
    \item \textbf{Generalised hypergeometric ensemble multilayer network regression (gHypE)}: This recent random graph approach~\cite{casiraghi2017multiplex} provides a statistical ensemble of all possible flow networks under the constraints of preserving in- and out-flows from each node, as well as respecting pairwise flow propensities of nodes. The multilayer network regression considers these propensities as latent variables, inferred from the edge features describing the dyadic relations between city locations. As opposed to conventional regression methods, this method intrinsically respects the network constraints.
    \item \textbf{Random Forest regression (RF)}: We follow the approach proposed in~\cite{spadon2019reconstructing} aimed at predicting inter-city mobility flows with a set of attributes describing each city. We adapt the same approach to our problem of intra-city flow prediction.  Following the described method, we use a Random Forest approach with eXtreme Gradient Boosting (\textit{XGBoost})~\cite{chen2016xgboost}  through 5-fold cross-validation, model and feature selection, and hyperparameter tuning. 
\end{itemize}

\subsection{Experimental setup}\label{sec:exp_setup}

For training and evaluating the three proposed models, we divide the dataset into a training, validation, and test set of edges. The subsets contain 70\%, 10\%, and 20\% of the edges respectively.
To construct the test set, we randomly select nodes in the graph and add their incident edges to the test set.
We ensure that an equal number of edges fall in each of the four bins split by flow magnitude. Hence, once a bin is full, no more edges are added to the test set that would fall into this bin.
We use the same procedure to construct the validation set. Nodes in the validation and test set are considered nodes of interest, while nodes in the training set are considered regular nodes. %The flow between nodes and interest and regular nodes is approximated using Equation~\ref{eq:approx_flows}.

We train all models on the same training set. To address the imbalance between flows of different magnitude, we resample the data such that each bin contains the same number of samples. We perform hyperparameter search to determine the optimal dimension of the intermediate representations, i.e. the outputs of the GCN and fully connected layers, the dropout rate, and the number of fully connected and GCN layers.
We select models based on the bin mean MAE (see Equation~\ref{eq:binmean}) achieved on the validation set. 
The selected models have a total of four fully connected layers. The GNN-based models use a single GCN layer.
We use a dimensionality of $32$ for intermediate representations and the dropout rate is set to $0.5$. 

We train for a total of 110 epochs using the Adam optimiser~\cite{kingma2015adam} with a batch size of 256 and a learning rate of $0.01$.
We reduce the learning rate by a factor of ten after 50 epochs and every 15 epochs after that.
We stop training early once the performance of the model does no longer improve in terms of bin mean MAE on the validation set.

We have also experimented with using different types of graph neural network layers including GAT layers~\cite{velickovic2018gat}, GIN layers~\cite{xu2018gin}, and Jumping Knowledge layers~\cite{xu2018jk}. We did not find these layers to improve performance on the validation data set and hence preferred the conceptually simpler GCN layers.

\section{Results}

\begin{figure*}
% \begin{multicols}{2}
    \begin{subfigure}{0.4078\textwidth}
    \includegraphics[width=\textwidth]{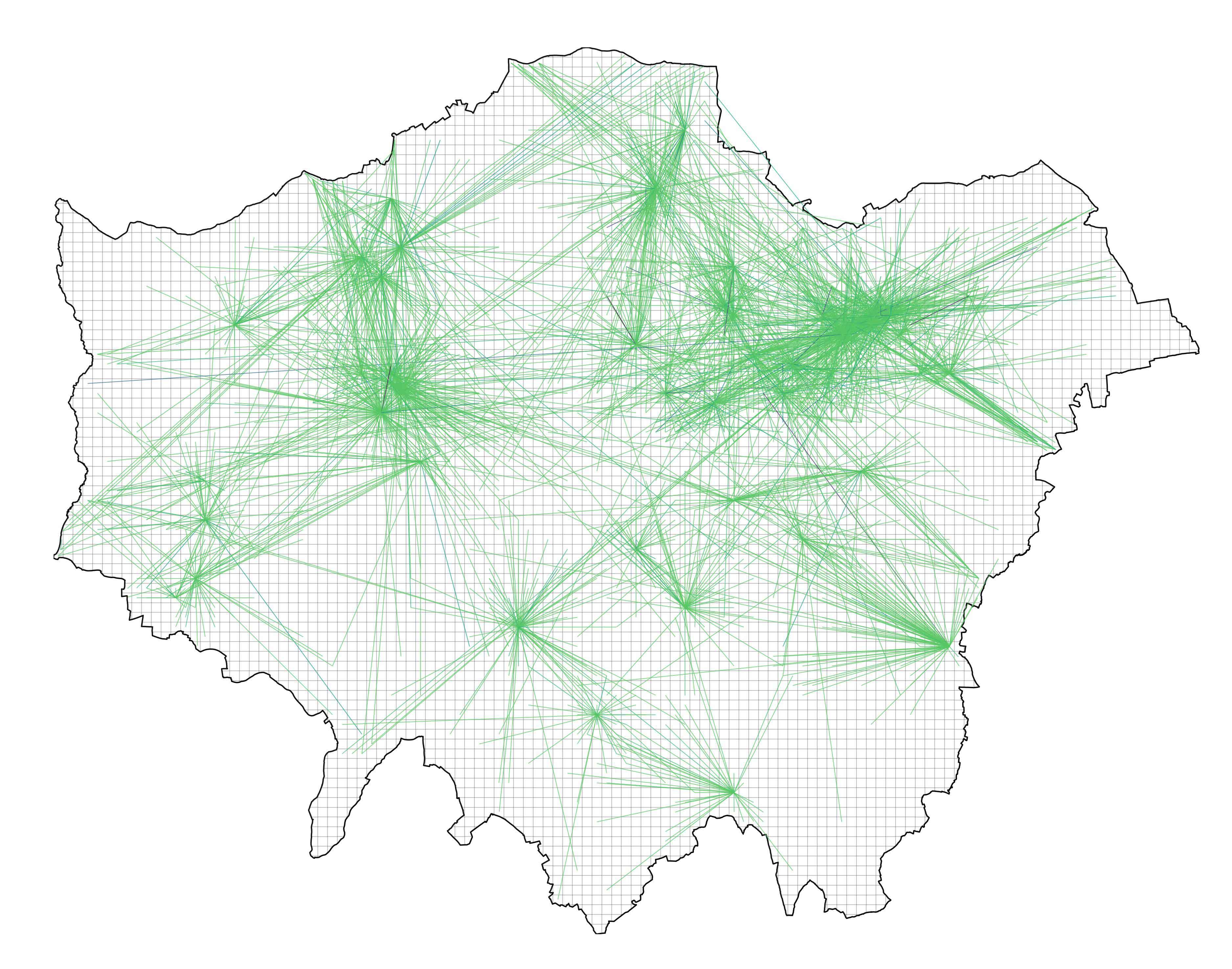} 
    \caption{}
    \end{subfigure}
    \hspace{0.8cm}
    \begin{subfigure}{0.4485\textwidth}
    \includegraphics[width=\textwidth]{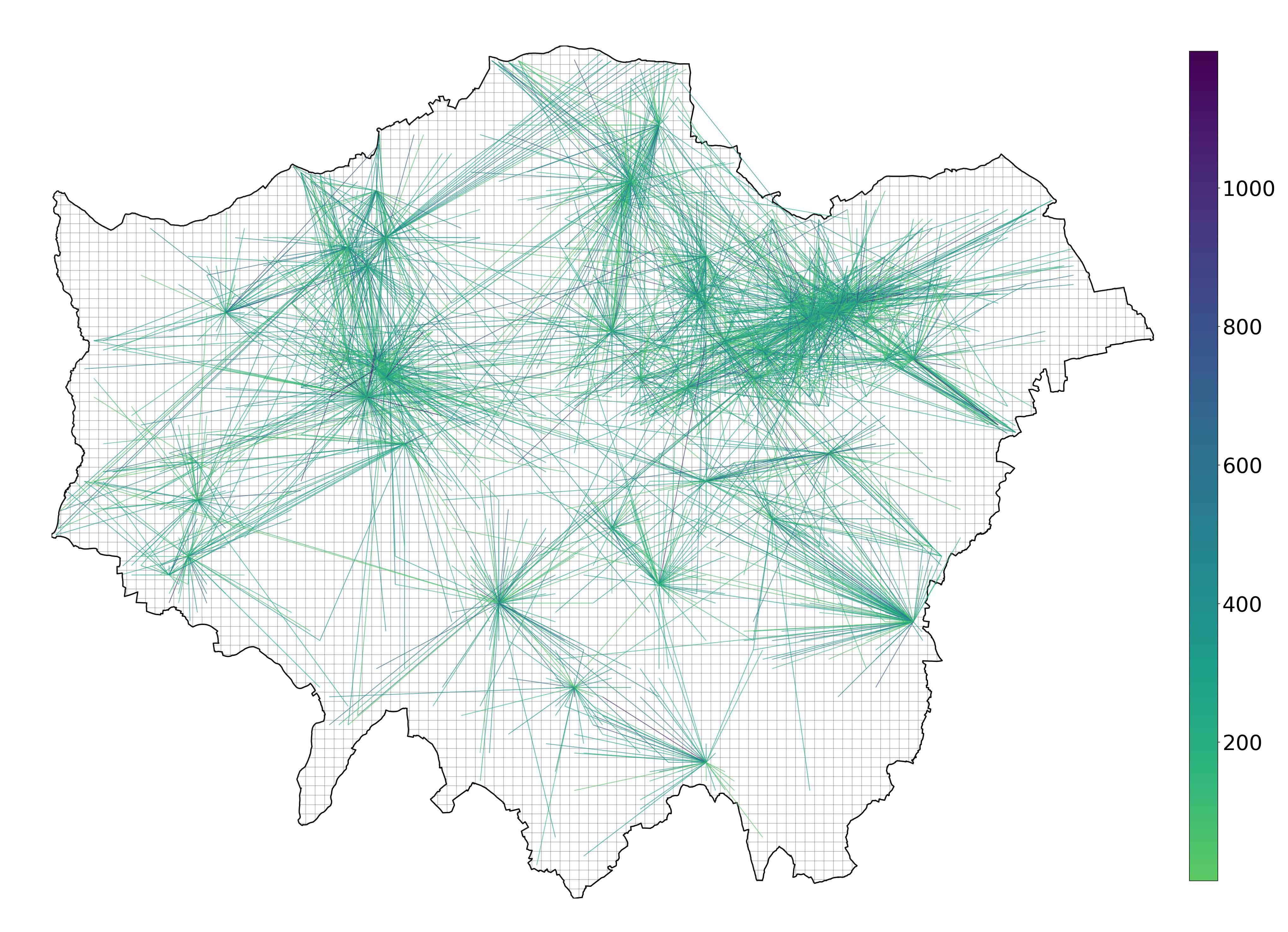}
    \caption{}
    \end{subfigure}
% \end{multicols}
\caption{MAE residuals of flows associated with test nodes (a) \textit{GNN-geo}. (b) \textit{XGBoost}.}
\label{resids}
\end{figure*}

\begin{table*}
    \centering
    \begin{tabular}{lcccccc}
    \toprule
    \textbf{MAE} & \textbf{Total} & $\bm{[0; 10)}$ & $\bm{[10; 10^2)}$ & $\bm{[10^2; 10^3)}$ & $\bm{[10^3; 10^4)}$ & \textbf{bin mean} \\
    \midrule
    \textbf{DC-GM} & $167.58$ & $64.88$ & $170.45$ & $881.98$ & $2176.35$ & $823.42$ \\
    \textbf{Huff} & $122.89$ & $48.21$ & $99.86$ & $511.41$ & $1476.72$ & $534.05$ \\
    \textbf{Poisson} & $106.74$ & $40.69$ & $88.56$ & $475.23$ & $1261.41$ & $466.47$ \\
    \textbf{NB} & $92.62$ & $33.02$ & $76.96$ & $431.44$ & $1087.12$ & $407.14$ \\
    \textbf{SAM} & $75.09$ & $19.31$ & $61.53$ & $395.01$ & $989.30$ & $366.29$ \\
    \textbf{gHypE} & $58.11$ & $9.02$ & $53.10$ & $346.96$ & $832.26$ & $310.34$ \\
    \textbf{XGBoost} & $31.59 \pm 5.88$ & $2.61 \pm 0.89$ & $45.12 \pm 11.06$ & $228.96 \pm 39.96$ & $549.83 \pm 84.79$ & $206.63 \pm 34.18$ \\
    \midrule
    \textbf{FCNN} & $12.55 \pm 0.91$ & $0.33 \pm 0.08$ & $28.97 \pm 4.93$ & $161.12 \pm 22.36$ & $408.88 \pm 36.59$ & $149.82 \pm 13.65$ \\
    \textbf{GNN-geo} & $13.34 \pm 2.51$ & $0.52 \pm 0.40$ & $31.63 \pm 9.68$ & $161.32 \pm 9.09$ & $422.04 \pm 25.70$ & $153.88 \pm 9.74$ \\
    \textbf{GNN-flow} & $15.35 \pm 4.23$ & $0.63 \pm 0.62$ & $38.66 \pm 16.65$ & $170.06 \pm 17.41$ & $458.05 \pm 64.56$ & $166.85 \pm 16.39$ \\
    \bottomrule
    \end{tabular}
\caption{Comparison of model performance in terms of mean absolute error grouped by flow magnitude.}
\label{MAE_table}
\end{table*}

\begin{table}
    \centering
    \resizebox{\linewidth}{!}{
    \begin{tabular}{lcccc}
    \toprule
     & \multirow{2}{*}{\textbf{SSI}} & \textbf{MAPE} & \multirow{2}{*}{\textbf{CPL}} & \multirow{2}{*}{\textbf{CPC}} \\
     & & $\bd{[10^3; 10^4)}$ & & \\
    \midrule
    \textbf{DC-GM} & $0.39$ & $162.59$ & $0.38$ & $0.49$ \\
    \textbf{Huff} & $0.48$ & $106.91$ & $0.56$ & $0.54$ \\
    \textbf{Poisson} & $0.46$ & $102.10$ & $0.57$ & $0.54$ \\
    \textbf{NB} & $0.54$ & $91.03$ & $0.62$ & $0.56$ \\
    \textbf{SAM} & $0.59$ & $66.65$ & $0.68$ & $0.58$ \\
    \textbf{gHypE} & $0.62$ & $52.99$ & $0.79$ & $0.60$ \\
    \textbf{XGBoost} & $0.67 \pm 0.02$ & $40.90 \pm 5.85$ & $0.86 \pm 0.02$ & $0.61 \pm 0.01$ \\
    \midrule
    \textbf{FCNN} & $0.71 \pm 0.00$ & $27.16 \pm 2.23$ & $1.0 \pm 0.00$ & $0.69 \pm 0.01$ \\
    \textbf{GNN-geo} & $0.70 \pm 0.01$ & $27.06 \pm 1.65$ & $1.0 \pm 0.00$ & $0.68 \pm 0.04$ \\
    \textbf{GNN-flow} & $0.71 \pm 0.02$ & $30.67 \pm 4.18$ & $1.0 \pm 0.01$ & $0.65 \pm 0.05$ \\
    \bottomrule
    \end{tabular}}
\caption{Comparison of model performance in terms of MAPE, SSI, CPL, and CPC.}
\label{SSI_table}
\end{table}

We compare our models to the baseline ones in terms of MAE in Table~\ref{MAE_table}. We find that all three neural network models outperform all the spatial interaction models (\textit{DC-GM}, \textit{Huff}, \textit{Poisson}, \textit{NB}, \textit{SAM}) as well as \textit{gHypE} and \textit{XGBoost} in terms of total MAE by a large margin. 
Crucially, the MAEs per bin reveal that the neural network models achieve high accuracy across bins relative to the magnitude of flows, hence the neural network does not only perform well on small flows, which are highly overrepresented in the dataset.

We also observe that there is no clear difference in the performance between the three neural network based models. Surprisingly, the graph neural networks (\textit{GNN-geo}, \textit{GNN-flow}) do not outperform the fully connected neural network \textit{FCNN}. This indicates that node neighbourhood information does not result in stronger predictive performance for this dataset and prediction task. We stress, however, that while \textit{FCNN} is not a graph neural network, it does use graph structural information by concatenating edge features with features of incident nodes. Furthermore, previous work on mobility flow prediction has omitted an explicit comparison of GNNs to fully connected neural networks, hence it remains unclear whether GNNs offer a predictive advantage in the urban mobility setting.

Finally, we compare the neural network models to the baselines in terms of SSI, MAPE of the largest bin, CPL, and CPC. These results also confirm that the neural network models find a better fit to the data compared to the state-of-the-art.

To further illustrate the effectiveness of the GNN models, we represent the MAE residuals on the London diagrammatic maps in Figure~\ref{resids}. These representations show the difference between predicted and ground-truth flows between the locations in the test set. We compare the state-of-the-art \textit{XGBoost} model with our \textit{GNN-flow} model and observe that the latter results in spatially smoother residuals.

% \begin{figure*}
%   \begin{subfigure}{0.46\textwidth}
%     \includegraphics[\linewidth]{London_gnn_resids.jpg}
%     \caption{}
%     \label{fig:resids1}
%   \end{subfigure}
%   \hfill
%   \begin{subfigure}{0.46\textwidth}
%     \includegraphics[width=\linewidth]{London_xgboost_resids.jpg}
%     \caption{}
%     \label{fig:resids2}
%   \end{subfigure}
%   \caption{MAE Residuals (a) \textit{GNN-geo}. (b) XGBoost.}
% \end{figure*}

% However, we find that the GNN models, despite their advantage of treating the flow network explicitly as a graph and pooling information from topological and geographical neighbourhoods when learning edge representations, do not outperform the simpler \textit{FCNN} model on our task. We note, however, that the mentioned advantage refers only to the convolution operation, and not to using graph information in general, since the \textit{FCNN} model, making use of node features incident to the edge in question, implicitly utilises graph information. 
% In connection with this, we raise concern about the results of the studies mentioned in Section~\ref{related_work} using graph neural networks for predicting mobility flows, as they do not compare their GNN models to the simpler \textit{FCNN} model we have proposed.  

\section{Conclusion}
% In this paper, we studied a problem of interest to urban planners, real estate developers, and other stakeholders involved in urban development projects.
In this paper, we formulated and addressed the problem of learning urban mobility flows between a location of interest and every other location in the city, given the array of socio-economic and structural features describing each location and the pairwise dyadic relations between them. 
We proposed three novel neural network architectures, using fully connected and graph convolutional layers, and compared them to a set of strong baseline models. We find that the neural network models achieve state-of-the-art performance and outperform the baselines by a large margin.
% However, we did not find the graph neural network models to outperform the simpler fully connected neural network model. 

% This is a finding worth investigating further, since graph neural networks have shown convincing improvements over conventional deep learning approaches in other fields, such as computational chemistry and natural language processing~\cite{zhang2019graph}. 
% We plan to study this further by first assessing the impact of grid resolution on model performance. 
% A well-known problem in spatial statistics, the scale of the aggregation unit may be a source of bias, particularly influencing the information contained in geographical neighbourhoods, which the \textit{GNN} attempts to capture. 
% Although some initial experiments with grid resolutions ranging from $500 \times 500$m to $1500 \times 1500$m cells showed no significant differences in \textit{GNN} model performance, further studies in this direction are required. 
% As a further study on the possibilities and limitations of GNNs on the formulated task, we also plan to test the ability of graph convolutions to capture topological and geographical neighbourhood information from flow data generated by our baseline models. We expect these parametric simulations to offer wider control over the data-generating process and help to better understand the findings of this paper.

In fulfilment of the stated objective, our work has direct utility to urban planners and policy makers in offering a technique for assessing mobility flows between an urban development project location and other locations in the city.

\section*{Acknowledgment}

F.L.O acknowledges funding from the Huawei Hisilicon Studentship at the Department of Computer Science and Technology of the University of Cambridge.
This work is partially funded by the EU H2020 programme under Grant Agreement No. 780754, ``Track \& Know''.

\bibliographystyle{ieeetr}
\bibliography{ref}
\end{document}

%% file: loglogplot.tex
% This file was created by tikzplotlib v0.9.1.
\begin{tikzpicture}

\definecolor{color0}{rgb}{1,0.270588235294118,0}

\begin{axis}[
axis line style={white!41.1764705882353!black},
legend cell align={left},
legend style={fill opacity=0.8, draw opacity=1, text opacity=1, draw=white!80!black},
log basis x={10},
log basis y={10},
tick pos=both,
x grid style={white!69.0196078431373!black},
xlabel={Flow count},
xmajorgrids,
xmin=2.94513727219967, xmax=6905.23319032926,
xmode=log,
xtick style={color=white!41.1764705882353!black},
y grid style={white!69.0196078431373!black},
ylabel style={rotate=-90.0},
ylabel={\(\displaystyle p(x)\)},
ymajorgrids,
ymin=8.49217806510829e-08, ymax=0.268844784115058,
ymode=log,
ytick style={color=white!41.1764705882353!black},
width=0.95\linewidth,
height=0.7\linewidth
]
\addplot [only marks, mark=x, draw=white!41.1764705882353!black, fill=white!41.1764705882353!black, opacity=0.6, colormap/viridis]
table{%
x                      y
4.19407894736842 0.130276000361825
10.5822368421053 0.0103405661659884
16.9703947368421 0.00491562006290755
23.3585526315789 0.00228166132906159
29.7467105263158 0.00149299246600309
36.1348684210526 0.00120605585764137
42.5230263157895 0.000761819407266611
48.9111842105263 0.000699441883709715
55.2993421052632 0.000472170732663286
61.6875 0.000380502893697065
68.0756578947368 0.000382401340066188
74.4638157894737 0.000265782491677208
80.8519736842105 0.000281241269254352
87.2401315789474 0.00020177772837535
93.6282894736842 0.000166792073858656
100.016447368421 0.000162452767872089
106.404605263158 0.000136416931952689
112.792763157895 0.000125026253737952
119.180921052632 0.000122042980872188
125.569078947368 8.94981859729374e-05
131.957236842105 9.16678389662208e-05
138.345394736842 8.84133594762962e-05
144.733552631579 8.08195739998041e-05
151.121710526316 7.2954581899152e-05
157.509868421053 5.96654573152916e-05
163.898026315789 6.15639036844145e-05
170.286184210526 5.50549447045648e-05
176.674342105263 4.66475393555916e-05
183.0625 5.72245976978479e-05
189.450657894737 4.09522002482231e-05
195.838815789474 5.07156387179979e-05
202.226973684211 3.85113406307791e-05
208.615131578947 3.66128942616562e-05
215.003289473684 4.47490929864687e-05
221.391447368421 2.87479021610043e-05
227.779605263158 3.28160015234104e-05
234.167763157895 3.30872081475708e-05
240.555921052632 2.98327286576459e-05
246.944078947368 2.98327286576458e-05
253.332236842105 2.76630756643625e-05
259.720394736842 2.38661829261168e-05
266.108552631579 2.49510094227582e-05
272.496710526316 2.11541166845126e-05
278.884868421053 2.38661829261168e-05
285.273026315789 2.00692901878707e-05
291.661184210526 1.73572239462667e-05
298.049342105263 2.22389431811541e-05
304.4375 1.89844636912292e-05
310.825657894737 1.78996371945876e-05
317.213815789474 1.68148106979457e-05
323.601973684211 1.68148106979459e-05
329.990131578947 1.7628430570427e-05
336.378289473684 1.24755047113792e-05
342.766447368421 1.4916364328823e-05
349.154605263158 1.24755047113791e-05
355.542763157895 1.00346450939355e-05
361.930921052632 1.27467113355396e-05
368.319078947368 1.05770583422562e-05
374.707236842105 9.76343846977504e-06
381.095394736842 1.35603312080209e-05
387.483552631579 1.13906782147374e-05
393.871710526316 9.49223184561462e-06
400.259868421053 8.9498185972937e-06
406.648026315789 1.03058517180959e-05
413.036184210526 1.1661884838898e-05
419.424342105263 8.9498185972937e-06
425.8125 1.13906782147375e-05
432.200657894737 6.23775235568955e-06
438.588815789474 8.94981859729378e-06
444.976973684211 7.05137222817086e-06
451.365131578947 8.40740534897287e-06
457.753289473684 7.05137222817086e-06
464.141447368421 5.69533910736877e-06
470.529605263158 5.15292585904789e-06
476.917763157895 5.15292585904794e-06
483.305921052632 5.15292585904794e-06
489.694078947368 5.96654573152914e-06
496.082236842105 5.69533910736877e-06
502.470394736842 4.61051261072706e-06
508.858552631579 1.00346450939355e-05
515.246710526316 3.79689273824581e-06
521.634868421053 2.7120662416042e-06
528.023026315789 6.50895897984997e-06
534.411184210526 4.88171923488747e-06
540.799342105263 5.15292585904798e-06
547.1875 6.23775235568955e-06
553.575657894737 4.61051261072706e-06
559.963815789474 2.7120662416042e-06
566.351973684211 5.15292585904789e-06
572.740131578947 4.33930598656664e-06
579.128289473684 4.06809936240623e-06
585.516447368421 4.33930598656672e-06
591.904605263158 2.71206624160415e-06
598.292763157895 3.25447948992498e-06
604.680921052632 4.0680993624063e-06
611.069078947368 2.71206624160415e-06
617.457236842105 4.33930598656664e-06
623.845394736842 4.0680993624063e-06
630.233552631579 4.06809936240623e-06
636.621710526316 3.25447948992498e-06
643.009868421053 4.33930598656672e-06
649.398026315789 2.16965299328332e-06
655.786184210526 2.71206624160415e-06
662.174342105263 4.61051261072706e-06
668.5625 3.25447948992504e-06
674.950657894737 4.06809936240623e-06
681.338815789474 2.71206624160415e-06
687.726973684211 3.52568611408546e-06
694.115131578947 4.88171923488747e-06
700.503289473684 2.44085961744374e-06
706.891447368421 2.16965299328336e-06
713.279605263158 2.71206624160415e-06
719.667763157895 2.71206624160415e-06
726.055921052632 5.4241324832084e-06
732.444078947368 3.25447948992498e-06
738.832236842105 1.35603312080208e-06
745.220394736842 2.16965299328336e-06
751.608552631579 1.62723974496249e-06
757.996710526316 2.71206624160415e-06
764.384868421053 2.98327286576462e-06
770.773026315789 1.89844636912291e-06
777.161184210526 2.44085961744374e-06
783.549342105263 1.89844636912291e-06
789.9375 2.44085961744378e-06
796.325657894737 2.16965299328332e-06
802.713815789474 8.13619872481246e-07
809.101973684211 2.44085961744378e-06
815.490131578947 3.25447948992498e-06
821.878289473684 3.79689273824581e-06
828.266447368421 2.7120662416042e-06
834.654605263158 2.44085961744374e-06
841.042763157895 1.62723974496249e-06
847.430921052632 2.16965299328336e-06
853.819078947368 1.08482649664166e-06
860.207236842105 1.35603312080208e-06
866.595394736842 5.42413248320831e-07
872.983552631579 1.08482649664168e-06
879.371710526316 1.08482649664166e-06
885.759868421053 2.16965299328332e-06
892.14802631579 1.62723974496252e-06
898.536184210526 2.44085961744374e-06
904.924342105263 5.42413248320831e-07
911.3125 1.62723974496252e-06
917.700657894737 5.42413248320831e-07
924.088815789474 8.13619872481246e-07
930.476973684211 1.08482649664168e-06
936.865131578947 8.13619872481246e-07
943.253289473684 8.13619872481246e-07
949.641447368421 1.08482649664168e-06
956.029605263158 1.35603312080208e-06
962.417763157895 1.35603312080208e-06
968.805921052632 1.08482649664168e-06
975.194078947368 5.42413248320831e-07
981.582236842105 8.13619872481246e-07
987.970394736842 2.71206624160415e-07
994.358552631579 1.08482649664168e-06
1000.74671052632 8.13619872481246e-07
1007.13486842105 1.89844636912291e-06
1013.52302631579 1.08482649664168e-06
1019.91118421053 1.08482649664166e-06
1026.29934210526 1.08482649664166e-06
1032.6875 5.42413248320831e-07
1039.07565789474 5.42413248320831e-07
1045.46381578947 5.4241324832085e-07
1051.85197368421 8.13619872481246e-07
1058.24013157895 8.13619872481246e-07
1064.62828947368 2.71206624160415e-07
1071.01644736842 1.08482649664166e-06
1077.40460526316 1.08482649664166e-06
1083.79276315789 1.35603312080212e-06
1090.18092105263 1.08482649664166e-06
1096.56907894737 1.62723974496249e-06
1102.95723684211 1.35603312080208e-06
1109.34539473684 8.13619872481246e-07
1115.73355263158 5.42413248320831e-07
1122.12171052632 8.13619872481275e-07
1128.50986842105 2.71206624160415e-07
1134.89802631579 1.35603312080208e-06
1141.28618421053 1.08482649664166e-06
1147.67434210526 1.35603312080208e-06
1154.0625 1.08482649664166e-06
1160.45065789474 1.08482649664166e-06
1166.83881578947 1.35603312080212e-06
1173.22697368421 2.71206624160415e-07
1179.61513157895 1.08482649664166e-06
1186.00328947368 8.13619872481246e-07
1192.39144736842 1.35603312080208e-06
1198.77960526316 8.13619872481246e-07
1205.16776315789 5.4241324832085e-07
1211.55592105263 5.42413248320831e-07
1217.94407894737 8.13619872481246e-07
1224.33223684211 5.42413248320831e-07
1230.72039473684 5.42413248320831e-07
1243.49671052632 1.08482649664166e-06
1249.88486842105 1.0848264966417e-06
1269.04934210526 5.42413248320831e-07
1275.4375 2.71206624160415e-07
1281.82565789474 2.71206624160415e-07
1288.21381578947 8.13619872481275e-07
1294.60197368421 1.08482649664166e-06
1300.99013157895 1.35603312080208e-06
1307.37828947368 8.13619872481246e-07
1313.76644736842 5.42413248320831e-07
1326.5427631579 5.42413248320831e-07
1332.93092105263 2.71206624160425e-07
1339.31907894737 2.71206624160415e-07
1345.70723684211 2.71206624160415e-07
1352.09539473684 1.08482649664166e-06
1358.48355263158 2.71206624160415e-07
1371.25986842105 8.13619872481275e-07
1377.64802631579 2.71206624160415e-07
1384.03618421053 2.71206624160415e-07
1396.8125 5.42413248320831e-07
1403.20065789474 5.42413248320831e-07
1409.58881578947 5.4241324832085e-07
1415.97697368421 8.13619872481246e-07
1428.75328947368 1.08482649664166e-06
1454.30592105263 2.71206624160425e-07
1460.69407894737 5.42413248320831e-07
1467.08223684211 1.08482649664166e-06
1473.47039473684 2.71206624160415e-07
1479.85855263158 8.13619872481246e-07
1486.24671052632 1.08482649664166e-06
1492.63486842105 5.4241324832085e-07
1511.79934210526 2.71206624160415e-07
1518.1875 5.42413248320831e-07
1524.57565789474 2.71206624160415e-07
1530.96381578947 2.71206624160425e-07
1550.12828947368 5.42413248320831e-07
1556.51644736842 2.71206624160415e-07
1562.90460526316 2.71206624160415e-07
1569.2927631579 8.13619872481246e-07
1582.06907894737 2.71206624160415e-07
1594.84539473684 5.42413248320831e-07
1601.23355263158 2.71206624160415e-07
1614.00986842105 2.71206624160425e-07
1620.39802631579 2.71206624160415e-07
1633.17434210526 5.42413248320831e-07
1645.95065789474 2.71206624160415e-07
1658.72697368421 5.4241324832085e-07
1671.50328947368 2.71206624160415e-07
1677.89144736842 2.71206624160415e-07
1690.6677631579 5.42413248320831e-07
1697.05592105263 2.71206624160425e-07
1703.44407894737 2.71206624160415e-07
1709.83223684211 2.71206624160415e-07
1728.99671052632 2.71206624160415e-07
1735.38486842105 5.42413248320831e-07
1741.77302631579 2.71206624160425e-07
1748.16118421053 2.71206624160415e-07
1754.54934210526 2.71206624160415e-07
1773.71381578947 2.71206624160415e-07
1780.10197368421 2.71206624160425e-07
1837.59539473684 2.71206624160415e-07
1843.98355263158 2.71206624160415e-07
1856.75986842105 2.71206624160415e-07
1863.14802631579 2.71206624160425e-07
1901.47697368421 2.71206624160425e-07
1907.86513157895 2.71206624160415e-07
1914.25328947368 2.71206624160415e-07
1933.4177631579 2.71206624160415e-07
1958.97039473684 5.42413248320831e-07
1965.35855263158 2.71206624160415e-07
1971.74671052632 2.71206624160415e-07
1990.91118421053 2.71206624160415e-07
2016.46381578947 2.71206624160415e-07
2022.85197368421 2.71206624160425e-07
2035.62828947368 2.71206624160415e-07
2048.40460526316 2.71206624160425e-07
2061.18092105263 2.71206624160425e-07
2067.56907894737 2.71206624160406e-07
2073.95723684211 2.71206624160425e-07
2080.34539473684 2.71206624160406e-07
2099.50986842105 2.71206624160406e-07
2131.45065789474 2.71206624160406e-07
2240.04934210526 2.71206624160425e-07
2265.60197368421 2.71206624160406e-07
2284.76644736842 5.4241324832085e-07
2323.09539473684 2.71206624160406e-07
2591.39802631579 2.71206624160425e-07
2655.27960526316 2.71206624160406e-07
2687.22039473684 5.42413248320811e-07
2699.99671052632 2.71206624160425e-07
2725.54934210526 2.71206624160425e-07
2821.37171052632 2.71206624160425e-07
2827.75986842105 2.71206624160406e-07
2846.92434210526 2.71206624160425e-07
2904.4177631579 2.71206624160425e-07
3000.24013157895 2.71206624160425e-07
3294.09539473684 2.71206624160406e-07
3479.35197368421 2.71206624160425e-07
3894.58223684211 2.71206624160425e-07
4852.80592105263 2.71206624160425e-07
};
\addlegendentry{Flow data}
\addplot [line width=2.0pt, mygreen, dotted]
table {%
4.19407894736842 0.112381296363926
10.5822368421053 0.0162784732763615
16.9703947368421 0.00607326918060483
23.3585526315789 0.00311719269203122
29.7467105263158 0.0018818422036964
36.1348684210526 0.00125374563282016
42.5230263157895 0.000892530255491418
48.9111842105263 0.000666395364054618
55.2993421052632 0.000515750662022438
61.6875 0.000410513768024209
68.0756578947368 0.000334187940414656
74.4638157894737 0.000277123067520032
80.8519736842105 0.000233373598519684
87.2401315789474 0.000199116953122688
93.6282894736842 0.000171806283155676
100.016447368421 0.000149692551246386
106.404605263158 0.000131542919129832
112.792763157895 0.000116468496361921
119.180921052632 0.0001038155169405
125.569078947368 9.30946142405147e-05
131.957236842105 8.39336936853738e-05
138.345394736842 7.60458242546931e-05
144.733552631579 6.92069230945353e-05
151.121710526316 6.32399660191212e-05
157.509868421053 5.80036328014486e-05
163.898026315789 5.33840202684281e-05
170.286184210526 4.92885122393076e-05
176.674342105263 4.56411884984358e-05
183.0625 4.23793470187294e-05
189.450657894737 3.9450841622442e-05
195.838815789474 3.68120239125961e-05
202.226973684211 3.442613784232e-05
208.615131578947 3.22620567540567e-05
215.003289473684 3.02932819674921e-05
221.391447368421 2.84971428684638e-05
227.779605263158 2.68541535131502e-05
234.167763157895 2.53474917429811e-05
240.555921052632 2.39625748893623e-05
246.944078947368 2.26867121523355e-05
253.332236842105 2.15088182365412e-05
259.720394736842 2.04191762262369e-05
266.108552631579 1.9409240267685e-05
272.496710526316 1.84714706101745e-05
278.884868421053 1.75991950875931e-05
285.273026315789 1.67864923117155e-05
291.661184210526 1.60280927780842e-05
298.049342105263 1.53192948164812e-05
304.4375 1.46558928961696e-05
310.825657894737 1.40341162557779e-05
317.213815789474 1.34505761950567e-05
323.601973684211 1.29022206607503e-05
329.990131578947 1.23862949968551e-05
336.378289473684 1.19003079224405e-05
342.766447368421 1.14420019572331e-05
349.154605263158 1.10093276434984e-05
355.542763157895 1.06004210180639e-05
361.930921052632 1.02135838750676e-05
368.319078947368 9.84726643173495e-06
374.707236842105 9.50005206897693e-06
381.095394736842 9.17064386814e-06
387.483552631579 8.85785270660177e-06
393.871710526316 8.5605867095715e-06
400.259868421053 8.27784188458318e-06
406.648026315789 8.00869378972504e-06
413.036184210526 7.75229010740502e-06
419.424342105263 7.50784401304676e-06
425.8125 7.27462824306483e-06
432.200657894737 7.0519697792102e-06
438.588815789474 6.83924507726291e-06
444.976973684211 6.6358757773704e-06
451.365131578947 6.44132484133131e-06
457.753289473684 6.25509306900816e-06
464.141447368421 6.07671595198741e-06
470.529605263158 5.90576082773415e-06
476.917763157895 5.74182430192888e-06
483.305921052632 5.5845299105271e-06
489.694078947368 5.43352599643173e-06
496.082236842105 5.28848377858612e-06
502.470394736842 5.14909559384104e-06
508.858552631579 5.01507329417544e-06
515.246710526316 4.88614678379932e-06
521.634868421053 4.76206268237814e-06
528.023026315789 4.64258310212011e-06
534.411184210526 4.52748452779131e-06
540.799342105263 4.41655678988923e-06
547.1875 4.30960212223552e-06
553.575657894737 4.20643429615913e-06
559.963815789474 4.10687782424758e-06
566.351973684211 4.01076722735959e-06
572.740131578947 3.91794635922739e-06
579.128289473684 3.82826778354237e-06
585.516447368421 3.74159219892066e-06
591.904605263158 3.65778790759441e-06
598.292763157895 3.57673032407484e-06
604.680921052632 3.49830152039193e-06
611.069078947368 3.42238980483574e-06
617.457236842105 3.34888933141228e-06
623.845394736842 3.27769973748461e-06
630.233552631579 3.20872580730131e-06
636.621710526316 3.14187715932323e-06
643.009868421053 3.07706795544656e-06
649.398026315789 3.01421663038998e-06
655.786184210526 2.95324563966534e-06
662.174342105263 2.89408122469003e-06
668.5625 2.83665319372306e-06
674.950657894737 2.78089471742e-06
681.338815789474 2.726742137904e-06
687.726973684211 2.67413479034259e-06
694.115131578947 2.6230148361042e-06
700.503289473684 2.57332710664462e-06
706.891447368421 2.52501895734303e-06
713.279605263158 2.47804013057049e-06
719.667763157895 2.43234262733139e-06
726.055921052632 2.38788058687058e-06
732.444078947368 2.34461017368728e-06
738.832236842105 2.30248947143993e-06
745.220394736842 2.2614783832669e-06
751.608552631579 2.22153853808369e-06
757.996710526316 2.18263320245145e-06
764.384868421053 2.14472719764185e-06
770.773026315789 2.1077868215516e-06
777.161184210526 2.0717797751456e-06
783.549342105263 2.03667509313166e-06
789.9375 2.00244307859095e-06
796.325657894737 1.96905524130899e-06
802.713815789474 1.93648423956975e-06
809.101973684211 1.90470382519293e-06
815.490131578947 1.87368879160965e-06
821.878289473684 1.84341492478645e-06
828.266447368421 1.81385895682069e-06
834.654605263158 1.78499852204265e-06
841.042763157895 1.75681211547098e-06
847.430921052632 1.72927905347867e-06
853.819078947368 1.70237943653625e-06
860.207236842105 1.67609411390809e-06
866.595394736842 1.65040465018564e-06
872.983552631579 1.62529329354959e-06
879.371710526316 1.60074294565947e-06
885.759868421053 1.57673713307646e-06
892.14802631579 1.5532599801307e-06
898.536184210526 1.53029618315051e-06
904.924342105263 1.50783098597608e-06
911.3125 1.48585015668495e-06
917.700657894737 1.46433996546153e-06
924.088815789474 1.44328716354669e-06
930.476973684211 1.42267896320793e-06
936.865131578947 1.40250301867386e-06
943.253289473684 1.38274740798046e-06
949.641447368421 1.36340061567968e-06
956.029605263158 1.34445151636398e-06
962.417763157895 1.32588935896309e-06
968.805921052632 1.30770375177203e-06
975.194078947368 1.28988464817178e-06
981.582236842105 1.27242233300623e-06
987.970394736842 1.25530740958122e-06
994.358552631579 1.23853078725354e-06
1000.74671052632 1.22208366957942e-06
1007.13486842105 1.205957542994e-06
1013.52302631579 1.19014416599476e-06
1019.91118421053 1.17463555880347e-06
1026.29934210526 1.15942399348272e-06
1032.6875 1.1445019844842e-06
1039.07565789474 1.12986227960758e-06
1045.46381578947 1.11549785134953e-06
1051.85197368421 1.10140188862392e-06
1058.24013157895 1.0875677888351e-06
1064.62828947368 1.07398915028719e-06
1071.01644736842 1.06065976491317e-06
1077.40460526316 1.04757361130861e-06
1083.79276315789 1.03472484805544e-06
1090.18092105263 1.02210780732203e-06
1096.56907894737 1.00971698872687e-06
1102.95723684211 9.97547053453141e-07
1109.34539473684 9.85592818602862e-07
1115.73355263158 9.73849251779396e-07
1122.12171052632 9.62311465887864e-07
1128.50986842105 9.50974714143518e-07
1134.89802631579 9.39834385278661e-07
1141.28618421053 9.28885998939116e-07
1147.67434210526 9.18125201261767e-07
1154.0625 9.07547760625066e-07
1160.45065789474 8.9714956356486e-07
1166.83881578947 8.86926610848207e-07
1173.22697368421 8.76875013698249e-07
1179.61513157895 8.66990990163586e-07
1186.00328947368 8.5727086162582e-07
1192.39144736842 8.47711049439318e-07
1198.77960526316 8.38308071697546e-07
1205.16776315789 8.2905854012051e-07
1211.55592105263 8.19959157058186e-07
1217.94407894737 8.11006712605044e-07
1224.33223684211 8.02198081820941e-07
1230.72039473684 7.93530222054001e-07
1243.49671052632 7.76605041022453e-07
1249.88486842105 7.68342023144896e-07
1269.04934210526 7.44318603823245e-07
1275.4375 7.36557340267902e-07
1281.82565789474 7.28915178091019e-07
1288.21381578947 7.21389709631453e-07
1294.60197368421 7.13978587508378e-07
1300.99013157895 7.06679522824271e-07
1307.37828947368 6.99490283429939e-07
1313.76644736842 6.9240869224918e-07
1326.5427631579 6.78560011935492e-07
1332.93092105263 6.71788829726172e-07
1339.31907894737 6.65117106608568e-07
1345.70723684211 6.58542917671267e-07
1352.09539473684 6.52064384152657e-07
1358.48355263158 6.45679672123217e-07
1371.25986842105 6.33184593371552e-07
1377.64802631579 6.27070771692008e-07
1384.03618421053 6.21043859242375e-07
1396.8125 6.09244287560819e-07
1403.20065789474 6.0346848451436e-07
1409.58881578947 5.9777330101298e-07
1415.97697368421 5.92157254002782e-07
1428.75328947368 5.81156805342378e-07
1454.30592105263 5.60043933280148e-07
1460.69407894737 5.54943043638833e-07
1467.08223684211 5.49910569414508e-07
1473.47039473684 5.44945301047962e-07
1479.85855263158 5.40046055481478e-07
1486.24671052632 5.35211675467034e-07
1492.63486842105 5.30441028895438e-07
1511.79934210526 5.1650053230067e-07
1518.1875 5.11973978817837e-07
1524.57565789474 5.07505853211192e-07
1530.96381578947 5.03095161201466e-07
1550.12828947368 4.90198055391671e-07
1556.51644736842 4.86007566599078e-07
1562.90460526316 4.81869844270091e-07
1569.2927631579 4.77784012349634e-07
1582.06907894737 4.69764605220397e-07
1594.84539473684 4.61942689493024e-07
1601.23355263158 4.58103784613254e-07
1614.00986842105 4.5056620911654e-07
1620.39802631579 4.46866036269969e-07
1633.17434210526 4.39599276527974e-07
1645.95065789474 4.32505940161226e-07
1658.72697368421 4.25580589954471e-07
1671.50328947368 4.18817999249561e-07
1677.89144736842 4.15496160560264e-07
1690.6677631579 4.08968346016246e-07
1697.05592105263 4.057611849159e-07
1703.44407894737 4.02591083231193e-07
1709.83223684211 3.99457476152827e-07
1728.99671052632 3.90270133138495e-07
1735.38486842105 3.87277066305853e-07
1741.77302631579 3.84317825428836e-07
1748.16118421053 3.81391906258727e-07
1754.54934210526 3.78498813868685e-07
1773.71381578947 3.7001168366067e-07
1780.10197368421 3.67245128478219e-07
1837.59539473684 3.43666414676587e-07
1843.98355263158 3.41185691861142e-07
1856.75986842105 3.36303055971666e-07
1863.14802631579 3.33900408385001e-07
1901.47697368421 3.20003753601686e-07
1907.86513157895 3.17771039922707e-07
1914.25328947368 3.15561289692885e-07
1933.4177631579 3.09066729484784e-07
1958.97039473684 3.00710485376983e-07
1965.35855263158 2.98673651642018e-07
1971.74671052632 2.96657156972148e-07
1990.91118421053 2.90727049068752e-07
2016.46381578947 2.8308921667438e-07
2022.85197368421 2.81226143661125e-07
2035.62828947368 2.77553999428304e-07
2048.40460526316 2.73952332596965e-07
2061.18092105263 2.70419363512858e-07
2067.56907894737 2.68678099741317e-07
2073.95723684211 2.6695336811093e-07
2080.34539473684 2.65244961460605e-07
2099.50986842105 2.60215667738201e-07
2131.45065789474 2.52141584149901e-07
2240.04934210526 2.27294997345618e-07
2265.60197368421 2.21976212861239e-07
2284.76644736842 2.18107041704829e-07
2323.09539473684 2.10662143568974e-07
2591.39802631579 1.67685523198114e-07
};
\addlegendentry{$ \alpha = -2.088 $}
\end{axis}

\end{tikzpicture}

%% file: model_architecture.tex
\resizebox{0.75\textwidth}{!}{%
\begin{tikzpicture}

\node[circle, scale=1.0, color=black, fill=mygreen, draw=mygreen, line width=0.3mm] (vi) at (0.0, 1.0) {$v_i$};
\node[circle, scale=1.0, color=black, fill=mygreen, draw=mygreen, line width=0.3mm] (vj) at (0.0, -1.0) {$v_j$};
\draw[draw=mygreen, line width=0.3mm] (vi) -- (vj);

\node[circle, scale=1.5, color=black, fill=mygreen, draw=mygreen, line width=0.3mm, opacity=0.7] (vi1) at (-1.2, 1.8) {};
\node[circle, scale=1.5, color=black, fill=mygreen, draw=mygreen, line width=0.3mm, opacity=0.7] (vi2) at (0.1, 2.6) {};
\node[circle, scale=1.5, color=black, fill=mygreen, draw=mygreen, line width=0.3mm, opacity=0.7] (vi3) at (1.2, 1.9) {};
\draw[draw=mygreen, line width=0.3mm, opacity=0.7] (vi) -- (vi1);
\draw[draw=mygreen, line width=0.3mm, opacity=0.7] (vi) -- (vi2);
\draw[draw=mygreen, line width=0.3mm, opacity=0.7] (vi) -- (vi3);
\draw[draw=mygreen, line width=0.3mm, opacity=0.7] (vi1) -- (vi2);

\node[circle, scale=1.5, color=black, fill=mygreen, draw=mygreen, line width=0.3mm, opacity=0.7] (vj1) at (-1.2, -1.9) {};
\node[circle, scale=1.5, color=black, fill=mygreen, draw=mygreen, line width=0.3mm, opacity=0.7] (vj2) at (0.2, -2.7) {};
\node[circle, scale=1.5, color=black, fill=mygreen, draw=mygreen, line width=0.3mm, opacity=0.7] (vj3) at (1.0, -1.8) {};
\draw[draw=mygreen, line width=0.3mm, opacity=0.7] (vj) -- (vj1);
\draw[draw=mygreen, line width=0.3mm, opacity=0.7] (vj) -- (vj2);
\draw[draw=mygreen, line width=0.3mm, opacity=0.7] (vj) -- (vj3);

\node (xi) at (0.75, 1.0) {$\bd{x}_i^v$};
\node (xj) at (0.75, -1.0) {$\bd{x}_j^v$};
\node (xij) at (-0.25, 0.0) {$e_{ij}$};
\node (xij) at (0.5, 0.0) {$\bd{x}_{ij}^e$};

\draw [decorate,decoration={brace,amplitude=10pt},xshift=-4pt,yshift=0pt] (1.7, -0.6) -- (1.7, -3.0);
\draw [decorate,decoration={brace,amplitude=10pt},xshift=-4pt,yshift=0pt] (1.7, 3.0) -- (1.7, 0.6);

\node[draw, line width=0.3mm, rounded corners=0.5mm] (concat) at (2.5, 0.0) {$\bd{\vert\vert}$};
\draw[->, line width=0.3mm] (xi) -- (concat);
\draw[->, line width=0.3mm] (xj) -- (concat);
\draw[->, line width=0.3mm] (xij) -- (concat);

\node[draw, line width=0.3mm, rounded corners=0.5mm] (FCN) at (4.5, 0.0) {FCN};
\draw[->, line width=0.3mm] (concat) -- node[above] {$\bd{\bar{x}}$} (FCN);

\node[draw, line width=0.3mm, rounded corners=0.5mm] (GCN1) at (4.5, 1.8) {GCN};
\draw[->, line width=0.3mm] (2.0, 1.8) -- (GCN1);
\node[draw, line width=0.3mm, rounded corners=0.5mm] (GCN2) at (4.5, -1.8) {GCN};
\draw[->, line width=0.3mm] (2.0, -1.8) -- (GCN2);

\node[draw, line width=0.3mm, rounded corners=0.5mm] (plus) at (7.0, 0.0) {$\bd{+}$};
\draw[->, line width=0.3mm] (GCN1) -- node[above] {$\cdot \phi_2$} (plus);
\draw[->, line width=0.3mm] (FCN) -- node[above] {$\cdot \phi_1$} (plus);
\draw[->, line width=0.3mm] (GCN2) -- node[below] {$\cdot \phi_2$} (plus);

\node[draw, line width=0.3mm, rounded corners=0.5mm] (FCNs) at (9.0, 0.0) {FCNs};
\draw[->, line width=0.3mm] (plus) -- (FCNs);

\node (out) at (11.0, 0.0) {$\hat{y}_{ij}$};
\draw[->, line width=0.3mm] (FCNs) -- (out);

\end{tikzpicture}
}